\documentclass{aa}

\usepackage{txfonts}

\usepackage{graphicx}	
\usepackage{amsmath}	
\usepackage{amssymb}	
\usepackage{gensymb}
\usepackage{tabulary}
\usepackage{subfigure}
\usepackage{epstopdf}
\usepackage{natbib}
\usepackage{afterpage}
\usepackage{float}
\usepackage{mathtools}
\usepackage{booktabs}
\usepackage{caption}
\usepackage{fancyvrb}
\usepackage{multirow}
\usepackage{cases}
\usepackage{tikz}
\usepackage{bm}
\usepackage[ruled,vlined,linesnumbered]{algorithm2e}
\usepackage{hyperref}
\usepackage[capitalise,nameinlink]{cleveref}

\newenvironment{smatrix}
  {\left(\begin{smallmatrix}}
  {\end{smallmatrix}\right)}

\usepackage{xr}		
\usetikzlibrary{matrix,shapes,arrows,positioning}



\defcitealias{cappellari2016structure}{C16}
\defcitealias{eisenstein1998hop}{EH96}
\defcitealias{stoughton2002sloan}{S02}
\defcitealias{graham2019bcatalogue}{Paper II}
\defcitealias{graham2019cenvironment}{Paper III}
\defcitealias{graham2019dclusters}{Paper IV}

\DeclarePairedDelimiter\floor{\lfloor}{\rfloor}

\begin{document}


\title{A group finder algorithm optimised for the study of local galaxy environment}

\author{Mark T. Graham$^1$\thanks{E-mail: mark.graham@physics.ox.ac.uk},
Michele Cappellari$^1$,}
 \institute{
$^1$Sub-department of Astrophysics, Department of Physics, University of Oxford, Denys Wilkinson Building, Keble Road, Oxford, OX1 3RH, UK\\
}

\date{Accepted XXX. Received YYY; in original form ZZZ}




  \abstract
{The majority of galaxy group catalogues available in the literature use the popular friends-of-friends algorithm which links galaxies using a linking length. One potential drawback to this approach is that clusters of point can be link with thin bridges which may not be desirable. Furthermore, these algorithms are designed with large-scales galaxy surveys in mind rather than small-scale, local galaxy environments, where attention to detail is important.}
{Here we present a new simple group finder algorithm, \texttt{TD-ENCLOSER}, that finds the group that \textit{encloses} a target galaxy of interest.}
{\texttt{TD-ENCLOSER} is based on the kernel density estimation method which treats each galaxy, represented by a zero-dimensional particle, as a two-dimensional circular Gaussian. The algorithm assigns galaxies to peaks in the density field in order of density in descending order (``Top Down") so that galaxy groups ``grow" around the density peaks. Outliers in under-dense regions are prevented from joining groups by a specified hard threshold, while outliers at the group edges are clipped below a soft (blurred) interior density level.}
{The group assignments are largely insensitive to all free parameters apart from the hard density threshold and the kernel standard deviation, although this is a known feature of density-based group finder algorithms, and operates with a computing speed that increases linearly with the size of the input sample. In preparation for a companion paper, we also present a simple algorithm to select unique representative groups when duplicates occur.}
{\texttt{TD-ENCLOSER} produces results comparable to those from a widely used catalogue, as shown in a companion paper. A smoothing scale of 0.3 Mpc provides the most realistic group structure.}


\maketitle



%

\section{Introduction}
\label{sec:intro2}
It has been known since the first large-scale galaxy surveys that galaxies are not randomly distributed throughout the Universe, but are preferentially found in groups and clusters. This structure traces the underlying dark matter distribution which cannot be observed directly. Moreover, many galaxy properties depend on the local environment, including morphology and colour \citep{blanton2005environment, blanton2009environment}. Therefore, it is of great interest to produce accurate and reliable group catalogues of nearby galaxies with which to study the properties of galaxies as a function of environment. While the first spectroscopic galaxy survey (CfA1 Redshift Survey; \citealp{huchra1982groups, geller1983groups}) only used a single slit to obtain the redshift, more recent surveys use multi-slit or fibre-optic spectrographs to observe hundreds of thousands of galaxies with spectroscopy. Notable examples are the Two Degree Field Galaxy Redshift Survey (2dFGRS; \citealp{colless2001survey}), the Galaxy And Mass Assembly survey (GAMA; \citealp{driver2009gama,driver2011gama}) and the Sloan Digital Sky Survey (see \citealp{york2000sloan} for a technical summary, \citealp{gunn20062} for a summary of the SDSS telescope, \citealp{smee2013multi} for a description of the spectrographs and \citealp{blanton2017sloan} for a summary of SDSS-IV). Collectively, these surveys have provided the basis for studying galaxy environments across huge samples.\par
The availability of such large datasets allows the opportunity to produce galaxy group catalogues. Many catalogues have been produced by various research teams using data from one or more of these surveys. Although some catalogues have been based purely on data from the 2dFGRS \citep{merchan2002groups,eke2004groups,yang2005groups} or the GAMA survey (G$3$Cv7, \citealp{robotham2011groups}), the most productive survey for group catalogues has been the SDSS. Most significant data releases have been complemented by a group catalogue based on the spectroscopic sample, including DR2 
\citep{miller2005clusters}, DR3 
\citep{merchan2005groups}, DR4
(\citealp{yang2007groups}, updated to DR7), DR5 
\citep{tago2008groups}, DR7 
\citep{tago2010groups, munoz2012groups}, DR8 
\citep{tempel2012groups}, DR10 
\citep{tempel2014groups} and DR12
\citep{tempel2017groups}.\par
The power of these catalogues lies in their scope for studying galaxy properties across large samples to obtain powerful statistical results. However, they are almost always based on the friends-of-friends (FoF) method to assign galaxies to groups \citep{huchra1982groups, davis1985structure}. This simple method uses a linking metric to assign particles to halos and as such is a frequent choice for assigning galaxies to halos in dark matter simulations (\citealp{eke2004groups,tempel2016groups}; see \citealp{knebe2013structure} for a review). The linking metric is usually defined to be a constant fraction of the mean particle separation. For magnitude-limited surveys (like the SDSS spectroscopic survey), the linking length varies with $z$ to account for the change in the luminosity function with $z$ \citep{huchra1982groups}. One potential issue with the FoF method is that groups can end up being joined by thin bridges, which may not be desirable (or even harmful). \cite{yang2005groups} combined the FoF method with an iterative procedure that first estimates the location, mass and radius of dark matter halos based on the galaxy distribution, before assigning galaxies to those halos and recomputing the halos. \cite{miller2005clusters} used a spherical aperture and information about the galaxy colours to identify clusters based on the probability of obtaining the observed galaxy distribution randomly.\par
Another independent method for estimating the underlying probability density function of some discrete data is the kernel-density estimation (KDE) method \citep{parzen1962kde}. The premise behind this approach is that by replacing particles/galaxies of zero size by kernels of non-zero size, a continuous probability density function can be obtained across the coordinate space. To find clumps or groups in the particle distribution, all one needs to do is locate local maxima in the density function. There are a number of algorithms which assign particles/galaxies to groups based on the density field, although the details of the method can vary somewhat between them. However, there are no group catalogues currently available that are based on redshift surveys and use the KDE method. See \cite{knebe2013structure} for a complete review of group-finder algorithms used in galaxy simulations.\par
Our ultimate goal is to study the environment of galaxies in the SDSS-IV Mapping Nearby Galaxies at Apache Point Observatory (MaNGA) survey \citep{bundy2015overview} in as much detail as possible. To this end, we have paid great attention to obtaining an accurate catalogue of galaxies and galaxy groups. A key advantage of the sample of galaxies observed by the MaNGA survey is that it is small enough that the neighbours of MaNGA galaxies can be assessed visually. We take the opportunity to develop an algorithm that assigns galaxies to groups based on the underlying galaxy distribution, with the intention of identifying groups that match what one might conclude from looking at the galaxy distribution by eye.\par
This paper is split as follows. In \cref{sec:td-encloser}, we present a new group-finder algorithm, \texttt{TD-ENCLOSER}\footnote{The name \texttt{TD-ENCLOSER} encapsulates the general idea behind our algorithm in that it considers galaxies by their density in descending order i.e. Top Down (TD). \texttt{ENCLOSER} refers to the fact that the algorithm finds the group that encloses a particular galaxy. As the algorithm was developed with a specific astronomical application in mind, we give it an unofficial acronym in the spirit of so many other acronyms in astronomy: Top Down-EfficieNt loCaL neighbOur SEarcheR.}, which adapts features of previous algorithms and is suitable for sample sizes of up to $\sim10^5$ particles. We focus on providing a simple routine that provides a careful treatment of small groups and cropping outliers from large groups. In \cref{sec:previous_KDE}, we summarise previous group finder algorithms. We use the ``hill-climbing'' method but only consider directions where a galaxy is known to exist. We also adapt parameters from the \texttt{HOP} method of \cite{eisenstein1998hop}, hereafter EH96. We then test our algorithm on a mock galaxy catalogue to demonstrate its operation and effectiveness (\cref{sec:test}).\par
In \cite{graham2019bcatalogue} (Paper II), we use \texttt{TD-ENCLOSER} to find the nearest neighbours to MaNGA galaxies. The set of neighbours that we find depends on the MaNGA galaxy, and if MaNGA galaxies are local to each other, then the same intrinsic groups may be found multiple times but with slight differences between each set. Ultimately we want to construct a group catalogue for MaNGA galaxies where each MaNGA galaxy lives in a well-defined environment. This requires us to select unique environments for each MaNGA galaxy, which we achieve in \cref{sec:multiplicity}. In \cite{graham2019cenvironment} (Paper III), we use this catalogue to conduct a large study of galaxy angular momentum and environment, with a few specific examples shown in \cite{graham2019dclusters} (Paper IV).\par
\section{KDE-based clustering}
\label{sec:group_finder}
\subsection{Previous KDE-based group finder algorithms}
\label{sec:previous_KDE}
There are many algorithms present in the literature that use a kernel density estimator to group particles into clusters. Many of these were optimised for $N$-body dark matter simulations and hence deal with $\mathcal{O}(10^6)$ particles. The first to be developed was \texttt{DENMAX} \citep{bertschinger1991simulations,gelb1994denmax} which uses an interpolation of the particle distribution to define a regular rectangular grid. Particles slide from their original locations towards a nearby dense grid cell with a force that is proportional to the local gradient (i.e. the particles follow a fluid equation). All particles that settle at the same peak are considered to be part of the same halo. Particles at the edges are clipped using an energy constraint (by comparing a particle's kinetic and potential energy at different timestamps). Spline Kernel Interpolative Denmax (\texttt{SKID}; \citealp{weinberg1997galaxy}) is an updated version of \texttt{DENMAX} that employs a spline kernel interpolation with a variable kernel size, rather than a regular grid of uniform kernel size. The densities are only measured at the particle locations, and particles then move in the same way as \texttt{DENMAX} towards density peaks. The \texttt{HOP} method (\citealp{eisenstein1998hop}; EH96) is inspired by \texttt{SKID} in that densities are only calculated at particle locations. However, instead of particles following the density field via a fluid equation, particles ``hop'' to the densest neighbour within the nearest $N_{\rm{hop}}$ neighbours. Particles hop until they reach the densest particle, and all particles that hop to the same particle are assigned to the same halo. A set of six parameters (not including the kernel bandwidth) are used to merge groups and clip outliers.\par
\texttt{DENCLUE} \citep{hinneburg1998efficient, hinneburg2007denclue} is another KDE-based group finder algorithm. From a given particle, the algorithm climbs hills defined by the density field, and assigns all particles that climb to the same hill to the same cluster. This method has the advantage of many density-based methods in that there is a unique result regardless of the order in which particles are considered. \texttt{DENCLUE 2.0} \citep{hinneburg2007denclue} includes a variable step size to reduce the number of iterations by considering the local gradients, without compromising on accuracy. It also has a noise threshold which is used to discount local maxima which fail to reach this threshold. However, the algorithm does not set a minimum threshold for particles to be considered as members of a cluster, and so a cluster can have members with density $\approx$ 0.\par
If limitations due to computing power or sample size were not an issue, then the precise and formal way to find groups would be to start from a particular galaxy and find the direction of maximal gradient in the density field. After moving a certain step size in that direction, the search would be repeated until the galaxy reaches a point where all gradients are negative. A helpful picture to have in mind is if the field were overturned so that peaks became valleys, then the galaxy would roll down in the direction of the steepest downwards slope, rather like a rain drop, before stopping at the bottom of the valley where all gradients are positive. We note that a similar method already exists and is known as ``mean-shifting'' or ``mode-seeking'' (e.g. \citealp{cheng1995mean}, see \citealp{carreira2015clustering} for a review). In this method, a kernel is placed over a point and is shifted towards the direction where the density (number of points) increases within the kernel (defined by the mean-shift vector). While the ``rain drop'' method would be the most rigorous solution to this problem, there are two drawbacks to implementing it computationally.\par
Firstly, the step size should be sensitive to the gradient so that a steeper gradient encourages a larger step size, as in \cite{hinneburg2007denclue}. This can be fairly straightforward to implement based on the equations of motion in a potential for example. However, this will be inefficient for points which are far away from the peak/valley, as the gradient will be small. The second obstacle is optimising the search for the direction of steepest gradient. Once the particle has \textit{initially} found this direction, the search can be limited to $\phi \pm \Delta \phi$, where $\phi$ is the current direction of the steepest gradient and $\Delta \phi$ is the field of view (width of an arc). While this works in principle, the path to the top of the peak/bottom of the valley has the potential to be much longer than the distance travelled as the crow flies, especially if the topology is complex.\par
One option to simplify this is to roll down \textit{from} peaks in the density and tag all particles that are met along the way. Here, the search stops at the foot of the hill where all gradients are positive. This approach requires prior knowledge of the location of the peaks but can, in theory, be more efficient than the method described above. Instead of moving from multiple points to a single location, this method moves outward from a single location assigning particles to the peak along the way. This ``hill-down'' approach was first applied in \texttt{HD-DENCLUE} by \cite{xie2007hill} with the intention of finding groups of connected points in medical imaging data (see also \citealp{xie2010density}). In their approach, the data are finely gridded and all points on the grid are added to the cluster with each successive step down the hill. The edge (foot) of the cluster is defined where the absolute value of the gradient falls below a predefined noise threshold. While this method works well for millions of particles (as is the case for imaging data), it becomes inefficient for smaller samples of a few hundred particles because all directions need to be searched from the point of view of the peak.\par
A similar approach was taken by \cite{springel2001clusters} who combined the FoF method with a ``top-down'' method that can identify the background density field and substructure in a dark matter simulation. Their algorithm, called \texttt{SUBFIND}, sorts particles by their density and then ``rebuilds'' the particle distribution by adding them to halos in order of decreasing density. Particles are only assigned to one subhalo so that they do not contribute to the mass of the parent halo, but \cite{springel2001clusters} find that this does not affect the parent halo a great deal as the substructure is usually at a scale that is small compared to the parent halo.\par

\subsection{A top-down approach to KDE-based clustering}
All of these algorithms have been optimised for millions of particles and hence are appropriate for producing group catalogues based on the dark matter distribution. However, we are interested in simply grouping galaxies together and obtaining directly observable relations and are not concerned with the dark matter distribution. As we are only focussing on the neighbours local to a specific sample of galaxies (the MaNGA galaxies), we do not need to consider large numbers of galaxies. Moreover, we would like to be able to detect all group sizes from two upwards. We would also like to be able to differentiate nearby peaks rather than merge them, which can happen with the \texttt{HOP} method for example (see fig. 1 of EH96).\par
Our approach is to combine a ``top-down'' method with a hill climbing method so that it is efficient for sample sizes of a few hundred to a thousand particles. To keep our algorithm as simple as possible, we will only consider straight lines between points, ignoring the surface topology. By considering galaxies in order of their density from highest to lowest, we can identify the peaks before attracting galaxies towards those peaks. Hence, rather than sliding or hopping from a particular galaxy, we take a ``top-down'' approach where we move out from regions of high density to low density.\par
\begin{algorithm*}
\footnotesize
\DontPrintSemicolon
\SetKwComment{Comment}{\# }{}
\KwData{$\{\bm{x},\bm{y}\}$ coordinates in physical units (Mpc) for $N$ galaxies \textit{relative} to the target galaxy at $(0,0)$}
\KwIn{$\sigma_{\rm{ker}}=0.3$ Mpc, $\rho_{\rm{outer}}=1.6$, $\rho_{\rm{saddle}}=4$, $\rho_{\rm{peak}}=4.8$, $N_{\rm{merge}}=4$ (defaults).}
\KwResult{$\{\bm{x},\bm{y}\}$ coordinates for $\mathcal{N}$ galaxies where $\mathcal{N}$ is the richness of the group that encloses the target galaxy.}
\SetKwData{mono}{mono}
Set $N$ to be the number of coordinate pairs.\; 
Construct coarse regular grid $C_g$.\;
\For{$i = 1$ to $N$}{
Add 2D Gaussian kernel of standard deviation $\sigma_{\rm{ker}}$, amplitude 1 and centre $(\bm{x}_i,\bm{y}_i)$ to $C_g$.\;}
Interpolate coarse grid $C_g$ using a 2D spline interpolator: $s = \texttt{spline}(C_g)$.\;
 \lIf(\tcp*[f]{\texttt{Target galaxy is isolated.}}){$s(0,0)<\rho_{\rm{outer}}$}{\textbf{return}}
\Else(\tcp*[f]{\texttt{Target galaxy must be in a group.}}){
	Evaluate $s$ at $\{\bm{x},\bm{y}\}$ to obtain density $\rho$ at each galaxy location: $\bm{\rho}=s(\bm{x},\bm{y})$.\;
	Remove all galaxies where $\rho < \rho_{\rm{outer}}$. \tcp*[f]{These galaxies are isolated.} \;
	Set $N$ to the number of (remaining) coordinate pairs.\;
	Set $\{\bm{P},\bm{x}^P,\bm{y}^P,\bm{\rho}^P\}$ to be an array of shape $(N,4)$ which will hold information about a galaxy's enclosing peak, namely $(x^P,y^P)$ coordinates and density $\rho^P$ of its central galaxy, and peak number $P$.\;
      Sort $\bm{x},\bm{y}$ and $\bm{\rho}$ by $\bm{\rho}$ in descending order such that $\bm{\rho}[1] > \bm{\rho}[2] > \bm{\rho}[3] \dots$.\;
   \Comment{First pass}
   Assign galaxy with density $\bm{\rho}[1]$ to peak 1: $\{\bm{P},\bm{x}^P,\bm{y}^P,\bm{\rho}^P\}[1,:] \longleftarrow (1,\bm{x}[1],\bm{y}[1],\bm{\rho}[1])$.\;
   Set $N_P$ to be the number of peaks: $N_P \longleftarrow 1$. \tcp*[f]{\texttt{Only 1 peak exists at this point.}}\;
   \For{$i = 2$ to $N$}{
   Select unique rows (peaks) from $\{\bm{P},\bm{x}^P,\bm{y}^P,\bm{\rho}^P\}$ where $\bm{P} \geq 1$: $\{\bm{P},\bm{x}^P,\bm{y}^P,\bm{\rho}^P\}_u \longleftarrow \{\bm{P},\bm{x}^P,\bm{y}^P,\bm{\rho}^P\}[\bm{P} \geq 1,:]$. \tcp*[f]{Only select identified peaks.}\;
     Sort $\{\bm{P},\bm{x}^P,\bm{y}^P,\bm{\rho}^P\}_{\rm{u}}$ in order of increasing distance from $(\bm{x}_{i},\bm{y}_{i})$.\;
       \While(\tcp*[f]{Only consider at most the 10 nearest peaks.}){$(j \leq N_P) \land (j \leq 10)$}{
       Check if the gradient of the connecting line between the galaxy at location $(\bm{x}_i,\bm{y}_i)$ and the peak at location $(\bm{x}_j^P, \bm{y}_j^P)$ changes sign or increases monotonically: $\mono \longleftarrow \texttt{MonotonicIncrease}(\bm{x}_i,\bm{y}_i,\bm{x}_j^P,\bm{y}_j^P,s,\rho_{\rm{cap}}=\bm{\rho}_j^P,\epsilon=-0.1)$.\;
      \If(\tcp*[f]{Assign galaxy to existing peak.}){\mono = True}{
    $\{ \bm{P},\bm{x}^P,\bm{y}^P,\bm{\rho}^P \}[i,:] \longleftarrow \{ \bm{P},\bm{x}^P,\bm{y}^P,\bm{\rho}^P \}_u[j,:]$; \textbf{break}}
    \Else($j \longleftarrow j + 1$ \tcp*[f]{\texttt{Move on to next peak.}}){}
      }
      \If(\tcp*[f]{\texttt{Galaxy cannot be added to any existing peak.}})
      {$\mono=\textrm{False}$}{Assign galaxy to be the central galaxy of a new peak: $\{ \bm{P},\bm{x}^P,\bm{y}^P,\bm{\rho}^P \}[i,:] \longleftarrow (N_P+1,\bm{x}[i],\bm{y}[i],\bm{\rho}[i])$, $N_P \longleftarrow N_P + 1$\;}}
  \Comment{Second pass}
    \For{$i = 1$ to $N$}{
     \If{$(\rho_{\rm{outer}} \leq \bm{\rho}_i < \rho_{\rm{saddle}}) \land (\bm{\rho}^P_i \geq \rho_{\rm{peak}})$}{
     Sort $\{ \bm{x},\bm{y},\bm{\rho} \}$ in order of increasing distance from $(\bm{x}_i,\bm{y}_i)$.\;
     Set $\rho_{\rm{max}}$ to be the maximum density of the nearest $N_{\rm{merge}}-1$ neighbours.\;
      \If{$(\bm{\rho}_i+\rho_{\rm{max}})/2 < \rho_{\rm{saddle}}$}{Eject galaxy from group: $\{ \bm{P},\bm{x}^P,\bm{y}^P,\bm{\rho}^P \}[i,:] \longleftarrow (0,\bm{x}[i],\bm{y}[i],\bm{\rho}[i])$}}
 
}
  \Comment{Third pass}
  Select galaxies that have been ejected from their original peaks in the SP: $\{ \bm{P},\bm{x}^P,\bm{y}^P,\bm{\rho}^P \}' \longleftarrow \{ \bm{P},\bm{x}^P,\bm{y}^P,\bm{\rho}^P \}[(\rho_{\rm{outer}} \leq \bm{\rho} < \rho_{\rm{saddle}}) \land (\bm{P} = 0),:]$, $\{ \bm{x},\bm{y},\bm{\rho} \}' \longleftarrow \{ \bm{x},\bm{y},\bm{\rho} \}[(\rho_{\rm{outer}} \leq \bm{\rho} < \rho_{\rm{saddle}}) \land (\bm{P} = 0),:]$.\;
Set $N'$ to be the number of rows (galaxies) in $\{ \bm{x},\bm{y},\bm{\rho} \}'$.\;
   Assign galaxy with density $\bm{\rho}'[1]$ to peak $N_P+1$: $\{ \bm{P},\bm{x}^P,\bm{y}^P,\bm{\rho}^P \}'[1,:] \longleftarrow \{ N_P,\bm{x}'[1],\bm{y}'[1],\bm{\rho}'[1] \}$.\;
   Set $N'_P$ to be the number of new peaks added in the TP: $N'_P \longleftarrow 1$.\;
  \For{$i = 2$ to $N'$}{
     Select unique rows (peaks) from $\{ \bm{P},\bm{x}^P,\bm{y}^P,\bm{\rho}^P \}'$ where $\bm{P} > N_P$: $\{ \bm{P},\bm{x}^P,\bm{y}^P,\bm{\rho}^P \}'_u \longleftarrow \{\bm{P},\bm{x}^P,\bm{y}^P,\bm{\rho}^P\}[\bm{P} > N_P,:]$. \tcp*[f]{Do not consider peaks from FP.}\;
     Sort $\{ \bm{P},\bm{x}^P,\bm{y}^P,\bm{\rho}^P \}'_u$ by distance from $(\bm {x}'_i, \bm {y}'_i)$.\;
     $\mono \longleftarrow \texttt{MonotonicIncrease}(\bm{x}_{i},\bm{y}_{i},\bm{x}_j^P,\bm{y}_j^P,s,\rho_{\rm{cap}}=0,\epsilon=-0.01)$\;
       \While{$(j \leq N'_P) \land (j \leq 10)$}{
      \lIf{\mono = True}{
    $\{ \bm{P},\bm{x}^P,\bm{y}^P,\bm{\rho}^P \}'[i,:] \longleftarrow \{ \bm{P},\bm{x}^P,\bm{y}^P,\bm{\rho}^P \}'_u[j,:]$; \textbf{break}.}
        \lElse($j \longleftarrow j + 1$){}
    }

      \lIf{$\mono=\textrm{False}$}{
      $\{ \bm{P},\bm{x}^P,\bm{y}^P,\bm{\rho}^P \}'[i,:] \longleftarrow (N'_P+1,\bm{x}'[i],\bm{y}'[i],\bm{\rho}'[i])$, $N'_P \longleftarrow N'_P + 1$}
      Select only galaxies which lie in the same peak as the target galaxy.\;
      }}
\caption{\textrm{\texttt{TD-ENCLOSER}}: Finds group enclosing a target galaxy.}
\label{fig:group_finder}
\end{algorithm*}

\begin{algorithm*}
\footnotesize
\DontPrintSemicolon
\SetKwComment{Comment}{\# }{}
\KwData{$x_0$, $y_0$, $x_1$, $y_1$, $s$}
\KwIn{$\rho_{\rm{cap}}=0$, $\epsilon=-0.1$}
\SetKwData{mono}{mono}
\KwResult{\mono}
\Comment{Sample line along intervals with spacing 1\% of its total length.}
Set $\bm{\rho}^{\rm{mid}}$ to be an array of length 101.\;
        \For{$i = 0$ to 100}{ 
          $x^{\rm{mid}} \longleftarrow x_0+(x_1-x_0)i/100$\;
          $y^{\rm{mid}}\longleftarrow y_0+(y_1-y_0)i/100$\;
        $\bm{\rho}^{\rm{mid}}_i \longleftarrow s(x^{\rm{mid}}, y^{\rm{mid}})$  \tcp*[f]{\texttt{Calculate density at the $i$th element.}}}
     \For{$i = 0$ to 99}{
     $\rho_{\rm{diff}} \longleftarrow \bm{\rho}_{\rm{mid},i+1}-\bm{\rho}_{\rm{mid},i}$\;
     \If{$[(\rho_{\rm{cap}} \geq 0) \land (\rho_{\rm{mid},i+1} \leq \rho_{\rm{cap}})] \lor (\rho_{\rm{cap}}=0)$}{
       \lIf( \tcp*[f]{\texttt{If gradient does not fall below $\epsilon$.}}){$\rho_{\rm{diff}} \geq \epsilon$}{$\mono \longleftarrow \textbf{True}$}
       \lElse{$\mono \longleftarrow \textbf{False}$; \textbf{break}}
  }}
  \KwRet $\mono$
  \;
  
\caption{\textrm{\texttt{MonotonicIncrease}}: Checks if the gradient of a line remains above a (small) threshold $\epsilon$ along the entire length of the line.}
\label{fig:monotonic}
\end{algorithm*}
\section{Description of \texttt{TD-ENCLOSER}}
\label{sec:td-encloser}
\subsection{Definition of algorithm parameters}
The algorithm we present here is similar to \texttt{SUBFIND} in that it considers galaxies by their density in decreasing order, but its function is an adaptation of the \texttt{HOP} method of \citetalias{eisenstein1998hop}. However, while there are some similarities, we have a different goal in mind to \citetalias{eisenstein1998hop}, namely that we want to identify small scale group environments in detail, while \texttt{HOP} was designed for $N$-body simulations of order 10$^6$ particles. \citetalias{eisenstein1998hop} solved two key problems regarding the separation of halos from their surroundings as well as the merging of groups by introducing six tunable parameters. Despite the added complexity, they showed that the result was insensitive to all but five of those parameters. We adopt four of these parameters, and adapt three of them to our specific requirements. Another similarity between our algorithm and the one of \citetalias{eisenstein1998hop} is that we make three passes of the data, although the manner in which our passes operate differ.\par
Now, we give details of the parameters that we adopt from \citetalias{eisenstein1998hop} (they used $\delta$ to denote density but we use $\rho$ instead, as $\delta$ can also represent a difference):
\begin{itemize}
\item $\rho_{\rm{outer}}$: This is the minimum density required for a galaxy to be considered as part of a group. By setting this parameter, \citetalias{eisenstein1998hop} prevent particles in underdense regions from joining groups. We retain this functionality of $\rho_{\rm{outer}}$ in this work. This is the only parameter found by \citetalias{eisenstein1998hop} to have a significant impact on the final group distribution.
\item $\rho_{\rm{saddle}}$: This is a second contour level which, if $\rho_{\rm{saddle}} \neq \rho_{\rm{outer}}$, can be used to separate two peaks which are joined by a thin bridge where $\rho \geq \rho_{\rm{outer}}$. It could also be used to join two peaks which are separated by a local minimum where $\rho \geq \rho_{\rm{saddle}}$. Our algorithm considers the density field rather than only the galaxies and so we do not need to separate nearby peaks within $\rho_{\rm{saddle}}$ as they will not be joined in the first place. However, we use $\rho_{\rm{saddle}}$ to exclude outliers from groups based on the density of one of their local neighbours.
\item $N_{\rm{merge}}$: In \citetalias{eisenstein1998hop}, this parameter is used to merge two nearby groups. If a particle and one of its nearest $N_{\rm{merge}}$ neighbours are in different groups, then a boundary pair is defined between the particle and the densest of the $N_{\rm{merge}}$ nearest neighbours. If the density of the boundary pair, defined to be the mean of the density of the two particles, is greater or equal to $\rho_{\rm{saddle}}$, then the two groups are merged. We use $N_{\rm{merge}}$ and $\rho_{\rm{saddle}}$ in a similar way to eject galaxies which lie far enough below $\rho_{\rm{saddle}}$. Specifically, we eject galaxies if the mean between their density and the maximum density of their $N_{\rm{merge}}-1$ neighbours is less than $\rho_{\rm{saddle}}$. Hence, this decision is determined by the local galaxy distribution and $\rho_{\rm{saddle}}$ is effectively a blurred boundary.
\item $\rho_{\rm{peak}}$: In \citetalias{eisenstein1998hop}, if the density $\rho$ of a peak is such that $\rho_{\rm{outer}} \leq \rho<\rho_{\rm{peak}}$, then the peak is only considered as part of a subgroup and is attached to a larger group with $\rho>\rho_{\rm{peak}}$. Here, we would like to detect small groups which may not have the density required to reach $\rho_{\rm{peak}}$. However, we do use $\rho_{\rm{peak}}$ to decide whether to disconnect outliers from groups using $\rho_{\rm{saddle}}$ and $N_{\rm{merge}}$.
\end{itemize}
The final parameter is the kernel size $\sigma_{\rm{ker}}$ although this parameter is general to all KDE-based methods and is not specific to \citetalias{eisenstein1998hop}.

\subsection{Algorithm methodology}
In what follows, we explain the methodology behind \texttt{TD-ENCLOSER}, referring to the pseudocode given in \cref{fig:group_finder} by line number (e.g. \texttt{line 1}) and the one-dimensional visualisation shown in \cref{fig:groupfinder_1d}. In \cref{fig:group_finder}, we only give the minimum amount of information required to implement the algorithm, leaving finer details to the text. To allow the reader to follow the decision making process, we have provided in \cref{tab:groupfinder_1d} values for position along the $x$-axis and $\rho$ (height) for each example galaxy in \cref{fig:groupfinder_1d}, as well as the group assignments at each pass. In our discussion, we use the default parameters which we introduce fully in \cref{sec:free_params}. These are: $\sigma_{\rm{ker}}=0.3$ Mpc, $\rho_{\rm{outer}}=1.6$, $\rho_{\rm{saddle}}=4$, $\rho_{\rm{peak}}=4.8$ and $N_{\rm{merge}}=4$. Finally, we use one-based indexing in \cref{fig:group_finder}, so that $\bm{x}[1]$ is the first element of the array $\bm{x}$.\par
\subsubsection{Setting up the grid}
The first step is to construct a two-dimensional coarse grid ($\bm{C}_g$) that covers the extent of the galaxy distribution (\texttt{line 2}). The spacing between grid elements is a compromise between resolution and computational power. To set the spacing, we assume that we only need a box 20 by 20 Mpc in size centred on the target galaxy (although there is no requirement that the box should be square). The spacing needs to be at the very least smaller than the kernel size which by default is 0.3 Mpc. However, it needs to be a small fraction of $\sigma_{\rm{ker}}$ because the density field should be independent of the position of the grid on the sky. If the spacing is equal to $\sigma_{\rm{ker}}$, then the result will differ dramatically if the grid becomes offset by $\sigma_{\rm{ker}}/2$ for example. We find that if the spacing is about $0.2\sigma_{\rm{ker}}$, then the contour morphology of the density field remains unchanged regardless of any offset. For simplicity, we choose the grid size to be $301^2$ giving a spacing of $20/300 \approx 0.0666$ Mpc. We choose 301 rather than 300 to ensure that a grid element is placed at the centre where the target galaxy lives.\par
\subsubsection{Calculating the density field}
We calculate the density field by placing two-dimensional circular Gaussians,

\begin{equation}
\hat f(x,y) = \exp\Bigg[ \frac12 [x-\bm{X},y-\bm{Y}]^T\cdot \Sigma^{-1}\cdot  (x-\bm{X},y-{Y}) \Bigg]\textrm{,}
\end{equation}
at each galaxy location $(x,y)$, where $\Sigma$ is the covariance matrix $\begin{smatrix} \sigma_{\rm{ker}}^2 & 0 \\
    0 & \sigma_{\rm{ker}}^2\end{smatrix}$, and $(\bm{X},\bm{Y})$ are the grid coordinates (\texttt{line 3}). The contribution from each kernel at each grid element is added to the density field (\texttt{line 4}). We do not use an adaptive kernel as this would produce unwanted substructure in dense groups (where the kernel is smaller to allow greater resolution) or group together galaxies which are isolated (because the kernel size increases in areas of lower density). The radial profile of the kernel does not have an effect on the end result and hence we choose to use a Gaussian kernel. We estimate the density field on a grid rather than the galaxy positions themselves to maintain a constant resolution across the field. We perform a two-dimensional interpolation of the density field\footnote{We use the \texttt{SciPy} implementation \texttt{RectBivariateSpline}.} (\texttt{line 5}), which allows us to calculate the density field at any location within the extent of the grid.\par
As with many group finder algorithms, there are potential edge effects being close to the boundary of the density field. It may be the case that the density at galaxies close to the edge will be underestimated, and so the algorithm may potentially miss groups at the boundary, leading to inconsistent clustering. However, as long as the boundary is at least twice the expected maximum group radius from a particular galaxy, then these effects will not affect the clustering near to the target galaxy.
\subsubsection{Checking for isolation}
\begin{figure*}
\centering
\includegraphics[width=\textwidth]{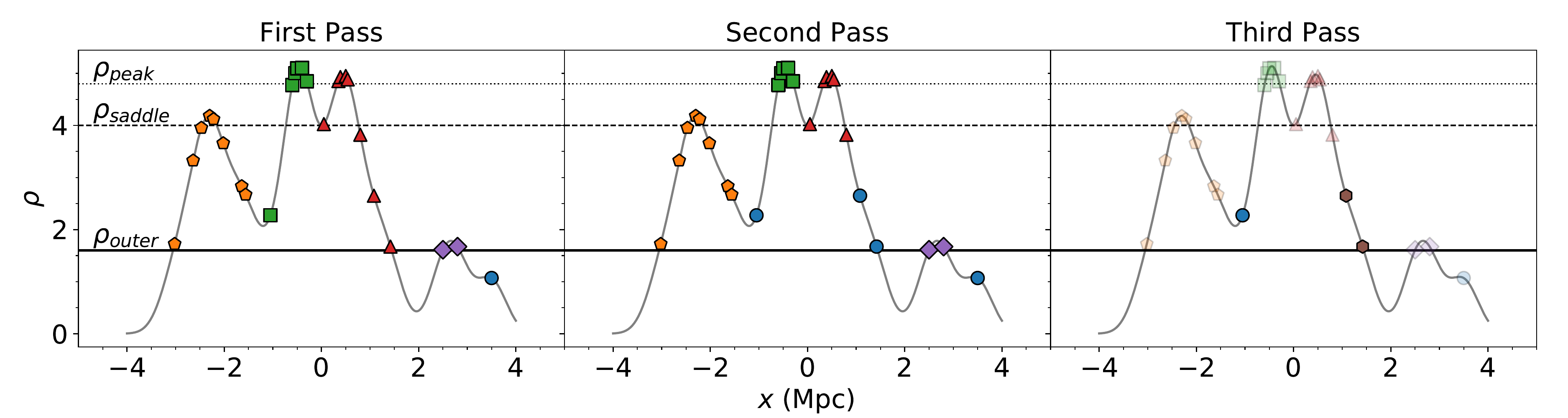}
\caption[One-dimensional visualisation of \texttt{TD-ENCLOSER}]{One-dimensional visualisation of the group finder algorithm \texttt{TD-ENCLOSER} introduced in this section. The points in the three panels are identical and each point represents a galaxy. Galaxies belonging to the same group as coloured accordingly. The contour levels from bottom up are $\rho_{\rm{outer}}$ (solid), $\rho_{\rm{saddle}}$ (dashed) and $\rho_{\rm{peak}}$ (dotted). The outcome of the first pass is shown in the first panel, where galaxies are assigned to groups based only on the contour morphology, shown as the grey curve. In the second pass, outliers are disconnected from their groups based on the the density of their $N_{\rm{merge}}-1$ neighbours and are tagged as isolated. In the third pass, each ejected galaxy is assigned to a new group. All galaxies which were not clipped in the second pass are shown as faint to indicate that they are considered to be absent in the third pass.}
\label{fig:groupfinder_1d}
\end{figure*}

Before running the main body of \texttt{TD-ENCLOSER}, we check to see if the target galaxy is isolated (\texttt{line 6}). If so, then the target galaxy cannot be part of a group, and so there is no need to proceed. Hence, \texttt{TD-ENCLOSER} terminates at this point, and the computing time is reduced. This would happen if the target galaxy were galaxy 25 in \cref{fig:groupfinder_1d}. If $\rho \geq \rho_{\rm{outer}}$ for the target galaxy, then \texttt{TD-ENCLOSER} proceeds as described below (\texttt{line 7}).\par

\subsubsection{Finding peaks in the first pass}
The first step in the first pass is to obtain the density at each galaxy by evaluating the spline interpolation at each galaxy location (\texttt{line 8}). We then automatically assign all isolated galaxies to be in groups of one member each (in essence removing them from the dataset; \texttt{line 9}). For each of the $N$ remaining galaxies with $\rho \geq \rho_{\rm{outer}}$ (\texttt{line 10}), we track which peak it belongs to as well as the position and density of the central galaxy (\texttt{line 11}). Before proceeding, we sort the $N$ galaxies by their density in descending order, so that we consider the galaxies at the peak densities first (Rank = 1, 2, ...) before adding nearby galaxies to the peaks (\texttt{line 12}). We then assign the galaxy at the densest peak to be the central galaxy of Group 1, as this is the only possible outcome for this galaxy (\texttt{line 13}). In our example, we assign galaxy 13 (Rank = 1) of density $\rho_{13}$ to be the central galaxy of Group 1. In this discussion, we use the term ``peak'' to describe a local maximum in the density field and the term ``group'' to refer to the galaxies that lie at a particular peak. There is a one-to-one correspondence between them i.e. Group 1 lies at peak 1.\par
We now enter a loop which loops over each of the remaining $N-1$ galaxies where $\rho_{\rm{outer}} \leq \rho < \rho_{13}$ (\texttt{line 15}). For each galaxy, we loop over all existing peaks (\texttt{line 16}) in order of increasing distance from the galaxy (\texttt{line 17}). We assume that a galaxy will not be able to join a more distant peak than the tenth nearest peak, and so if there are more than 10 peaks, we only consider the nearest 10 (\texttt{line 18}). 
The first galaxy we encounter in the loop in our example is galaxy 12 (Rank = 2). To decide whether or not to assign galaxy 12 to Group 1, which is the only existing group at this point, we check if the connecting line between galaxies 12 and 13 increases monotonically. Our assumption is that if it does increase monotonically in $\rho$, then galaxies 12 and 13 belong to the same group. However, we do require the whole length of the line to increase monotonically for the following reason. The local maxima in the density field will almost certainly not be at the location of any galaxy. However, we do not know the density field at ``every'' location within the grid and so we do not know the \textit{precise} locations of each local maxima. Instead, we only have knowledge of the galaxy that is closest to each local maxima by virtue of their high density $\rho$. Since we do not know in which direction the true local maxima lies with respect to the nearest galaxy, we have to treat the galaxy as if it is at the \textit{precise} location of the maxima. This means that when we move along the connecting line between two galaxies from low to high density (e.g. from galaxy 12 to galaxy 13), we stop when we reach the density of the upper galaxy (e.g. $\rho_{13}$) for the \textit{first} time.\par
We describe this check for monotonic increase in \cref{fig:monotonic}. First, we obtain the density $\rho$ at 101 equally spaced intervals along the line, where the spacing is equivalent to 1\% of the length of the line (\texttt{lines 1-5} of \cref{fig:monotonic}). We then iterate over each element starting from the low density end (i.e. the location of the galaxy of which we are deciding whether or not to assign to the peak; \texttt{line 6}). For each iteration, we calculate the difference between the element and the next one (\texttt{lines 7}). If the difference does not drop below $\epsilon=-0.1$, then the gradient of the line at the element is positive, and the algorithm moves on to the next element. We allow the gradient to go slightly negative (a difference of $\epsilon=-0.1$) to essentially account for noise in the density field. After some tests, we realised that some galaxies were being cut off from the peak even though the gradient was essentially flat to within 10\%. This could happen if a density contour happened to lie in parallel with the direction of the connecting line for example. We do not consider $\epsilon$ as a free parameter as it merely represents the uncertainty in the gradient. If the difference between each element does not fall below $\epsilon$ up until $\rho_{\rm{cap}}$ is reached, then we consider the two galaxies to be connected and part of the same peak (\texttt{lines 9-10}). On the other hand, if the difference does drop below $\epsilon$, then the two galaxies are not connected, and the main body of \texttt{TD-ENCLOSER} proceeds. In the following text, we represent these two outcomes with either $\texttt{MonotonicIncrease}(x_b,y_b,x_a,y_a,\texttt{spline}=s,\rho_{\rm{cap}}=\rho_a,\epsilon=-0.1)=True$ or $\texttt{MonotonicIncrease}(x_b,y_b,x_a,y_a,\texttt{spline}=s,\rho_{\rm{cap}}=\rho_a,\epsilon=-0.1)=False$ respectively, where $x_b$ and $y_b$ are the coordinates for galaxy $b$ (the low density galaxy).\par
In our example, we find that for galaxies 12 (Rank = 2) and 13 (Rank = 1), $\texttt{MonotonicIncrease}(x_{12},y_{12},x_{13},y_{13},\texttt{spline}=s,\rho_{\rm{cap}}=\rho_{13},\epsilon=-0.1)=True$ (\texttt{line 19}). The same is true for galaxy 11 (Rank = 3), and hence galaxies 12 and 11 are assigned to Group 1. However, $\texttt{MonotonicIncrease}(x_{18},y_{18},x_{13},y_{13},\texttt{spline}=s,\rho_{\rm{cap}}=\rho_{13},\epsilon=-0.1)=False$ (\texttt{line 23}), and hence galaxy 18 (Rank = 4) cannot be part of the same density peak as galaxy 13. Galaxy 18 is then designated as the central galaxy of Group 2 (\texttt{line 24}). We repeat this step until all galaxies above $\rho_{\rm{outer}}$ are assigned to groups/peaks (\texttt{line 15}).\par
\subsubsection{Ejecting outliers in the second pass}
Since the only requirement for assigning a galaxy to a peak in the first pass is that the connecting line between the galaxy and the peak increases monotonically up until a predetermined level, galaxies are assigned to peaks regardless of their distance from the peak. Hence, there may be outliers at the ``foot'' of the peak. Hence, we use a second pass to decide whether or not to eject those outliers. For each galaxy, we check if the density of the central galaxy of its enclosing group has reached $\rho_{\rm{peak}}$ (Groups 1 and 2 in our example) and if the density of the galaxy satisfies $\rho_{\rm{outer}} \leq \rho < \rho_{\rm{saddle}}$ (\texttt{line 26}). For each of these galaxies (\texttt{line 25}), we use the density of one of the nearest $N_{\rm{merge}}-1$ neighbours to decide if the galaxy should be removed from the group. Specifically, we take the maximum density from the nearest $N_{\rm{merge}}-1$ neighbours (\texttt{line 28}) and consider if the mean of this density and the density of the galaxy in question is less than $\rho_{\rm{saddle}}$ (\texttt{line 29}). If so, then the galaxy is clipped from the peak (\texttt{line 30}). Hence, $\rho_{\rm{saddle}}$ is not a rigid boundary but is in fact blurred according to the local galaxy distribution. By setting $\rho_{\rm{peak}}>\rho_{\rm{saddle}}$, we ensure that the second pass does not affect groups which just reach $\rho_{\rm{saddle}}$, and hence only groups with dense peaks will be clipped. For example, if $\rho_{\rm{peak}}=\rho_{\rm{saddle}}$, then Group 3 would be clipped. However, as the peak only just reaches $\rho_{\rm{saddle}}$, only the central three galaxies would remain unclipped and hence Group 3 would be split up even though it is a clearly defined peak.\par

\begin{table}
\centering
\caption{Table listing the relevant information about the example galaxies shown in \cref{fig:groupfinder_1d}. Galaxies are numbered from left to right as given in Column (1). Column (2) gives the position along the $x$ axis in Mpc and Column (3) gives the density $\delta$ at each galaxy. Column (4) lists the density rank for each galaxy with column (1) being the galaxy with the highest density. Columns (5), (6) and (7) give the group number that each galaxy belongs to after the first, second and third passes respectively. A value of zero indicates that a galaxy is isolated.}
\label{tab:groupfinder_1d}
\resizebox{0.47\textwidth}{!}{%
\begin{tabular}{@{}ccccccc@{}}
\toprule
\begin{tabular}[c]{@{}c@{}}Galaxy\\No.\\(1)\end{tabular} &
\begin{tabular}[c]{@{}c@{}}$x$\\(Mpc)\\(2)\end{tabular} &
\begin{tabular}[c]{@{}c@{}}$\rho$\\ \\(3)\end{tabular} &
\begin{tabular}[c]{@{}c@{}}Rank\\ \\(4)\end{tabular} &
\begin{tabular}[c]{@{}c@{}}Group No.\\(1$^{\rm{st}}$ pass)\\(5)\end{tabular} &
\begin{tabular}[c]{@{}c@{}}Group No.\\(2$^{\rm{nd}}$ pass)\\(6)\end{tabular} &
\begin{tabular}[c]{@{}c@{}}Group No.\\(3$^{\rm{rd}}$ pass)\\(7)\end{tabular} \\ \midrule
1 & -3.02 & 1.723 & 21 & 3 & 3 & 3\\
2 & -2.64 & 3.325 & 16 & 3 & 3 & 3\\
3 & -2.47 & 3.953 & 13 & 3 & 3 & 3\\
4 & -2.30 & 4.183 & 10 & 3 & 3 & 3\\
5 & -2.22 & 4.120 & 11 & 3 & 3 & 3\\
6 & -2.02 & 3.657 & 15 & 3 & 3 & 3\\
7 & -1.64 & 2.831 & 17 & 3 & 3 & 3\\
8 & -1.56 & 2.669 & 18 & 3 & 3 & 3\\
9 & -1.05 & 2.274 & 20 & 1 & 0 & 0\\
10 & -0.60 & 4.775 & 9 & 1 & 1 & 1\\
11 & -0.54 & 5.003 & 3 & 1 & 1 & 1\\
12 & -0.50 & 5.096 & 2 & 1 & 1 & 1\\
13 & -0.40 & 5.105 & 1 & 1 & 1 & 1\\
14 & -0.30 & 4.846 & 8 & 1 & 1 & 1\\
15 & 0.05 & 4.026 & 12 & 2 & 2 & 2\\
16 & 0.35 & 4.853 & 7 & 2 & 2 & 2\\
17 & 0.39 & 4.929 & 5 & 2 & 2 & 2\\
18 & 0.50 & 4.950 & 4 & 2 & 2 & 2\\
19 & 0.54 & 4.886 & 6 & 2 & 2 & 2\\
20 & 0.80 & 3.822 & 14 & 2 & 2 & 2\\
21 & 1.08 & 2.651 & 19 & 2 & 0 & 5\\
22 & 1.42 & 1.673 & 22 & 2 & 0 & 5\\
23 & 2.50 & 1.612 & 24 & 4 & 4 & 4\\
24 & 2.80 & 1.672 & 23 & 4 & 4 & 4\\
25 & 3.50 & 1.070 & 25 & 0 & 0 & 0\\
\bottomrule
\end{tabular}}
\end{table}

\subsubsection{Cleaning up in the third pass}
Once the second pass is completed, we use a third pass to attempt to group together the galaxies that were ejected during the second pass. If we do not have a ``clean up'' step, then we will be left with many isolated galaxies in potentially dense environments above $\rho_{\rm{outer}}$. We keep the method of the third pass identical to that of the first pass, but with three key differences. Firstly, we wish to prevent these galaxies from being reassigned to their original groups. Hence, we restrict the available neighbours to only those galaxies which have been ejected from groups in the second pass (\texttt{line 31}). Secondly, we lift the cap from the first pass that prevents galaxies on opposite sides of a peak from missing each other (\texttt{line 38}). In this case, we have the opposite situation to the first pass where we do \textit{not} want outliers forming very broad but low density peaks which encompass other (higher density) peaks. Hence, it is necessary to relax this rule here and require that the connecting line increases along its full length. Finally, we decrease the magnitude of the noise threshold by a factor of 10 to give $\epsilon=-0.01$ so that galaxies cannot form groups over long distances.\par
In our one-dimensional example, the first galaxy to be considered in the third pass is galaxy 21. As far as this galaxy is concerned, there are no other peaks and hence it becomes the central galaxy of a new group (\texttt{line 33}). Galaxy 9 is the next to be considered. For galaxy 9, $\texttt{MonotonicIncrease}(x_{9},y_{9},x_{21},y_{21},\texttt{spline}=s,\rho_{\rm{cap}}=\rho_{21},\epsilon=-0.01)=True$, but galaxies 9 and 21 are clearly not connected (see right panel of \cref{fig:groupfinder_1d}). Hence this is why the cap that is present in the first pass is lifted in the third pass, so that $\texttt{MonotonicIncrease}(x_{9},y_{9},x_{21},y_{21},\texttt{spline}=s,\rho_{\rm{cap}}=0,\epsilon=-0.01)=False$ (\texttt{line 42}). Thus, galaxy 9 becomes an isolated galaxy where $\rho > \rho_{\rm{outer}}$ (\texttt{line 38}). Finally, $\texttt{MonotonicIncrease}(x_{22},y_{22},x_{21},y_{21},\texttt{spline}=s,\rho_{\rm{cap}}=0,\epsilon=-0.01)=True$ (\texttt{line 40}). These two galaxies make up a galaxy pair where the separation is (marginally) larger than $\sigma_{\rm{ker}}$. This is indeed possible, but in practice, we find that the separation between a pair of galaxies is rarely larger than about $1.3\sigma_{\rm{ker}}$.\par

\subsection{Specifying default parameter values}
\label{sec:free_params}
Now we have described \texttt{TD-ENCLOSER}, we describe our best choices for the five free parameters. A common choice for the bandwidth, $\sigma_{\rm{ker}}$, is a ``rule-of-thumb'' estimation known as Scott's rule\footnote{BW=$n^{-1/6}$ for two-dimensional data where $n$ is the number of points.} (Scott 1992). However, this choice depends on the number of data points which is our case is not a constant. Furthermore, we wish to have a physically motivated value for $\sigma_{\rm{ker}}$. We first inspect by eye and deduce empirically which value best describes the galaxy distribution. For small values ($\sigma_{\rm{ker}} \sim 0.1$ Mpc), we find that galaxies struggle to form groups, and if large groups do form, then unphysical substructure appears. For large values ($\sigma_{\rm{ker}} \sim 0.5$ Mpc), then groups which are clearly distinct from each other merge to form larger groups which are not likely to be physically bound. We settle on a value of $\sigma_{\rm{ker}} = 0.3$ Mpc.\par
We follow \citetalias{eisenstein1998hop} by defining the three contour levels in terms of $\rho_{\rm{outer}}$. $\rho_{\rm{outer}}$ must be greater than one (in order to exclude isolated galaxies) and less than two (so that galaxy pairs are not missed)\footnote{$\rho_{\rm{outer}}$ only needs to be smaller than the minimum group richness that one wishes to be sensitive to.}. To set $\rho_{\rm{outer}}$, we consider a pair of galaxies separated by a distance $D$, one of which is the target galaxy. If $D=\sigma_{\rm{ker}}$, then the total density at the target galaxy is very close to 1.6\footnote{$e^0+e^{\frac{-(-\sigma_{\rm{ker}})^2}{2\sigma_{\rm{ker}}^2}}=1+e^{-0.5}=1.60653...$.}, regardless of the kernel size. Hence, we set $\rho_{\rm{outer}}=1.6$ so that we detect pairs of galaxies which are closer than the kernel size. This assumes that the pair is in isolation which is an unrealistic scenario. However, a Gaussian with amplitude 1 at a distance $D=3\sigma_{\rm{ker}}=0.9$ Mpc measures only $\sim0.01$ at $D=0$ and hence will only contribute about that much to the density at the pair\footnote{In any case, our approximation of 1.6 is about 0.6\% smaller than the exact value so these two effects essentially cancel out.}. Having defined the two parameters that will affect the group assignments the most, we use the default recommendations suggested by \citetalias{eisenstein1998hop} for the remaining three parameters. These are: $\rho_{\rm{saddle}}=2.5\rho_{\rm{outer}}$, $\rho_{\rm{peak}}=3\rho_{\rm{outer}}$ and $N_{\rm{merge}}=4$.

\begin{figure*}
\centering
\includegraphics[width=0.75\textwidth]{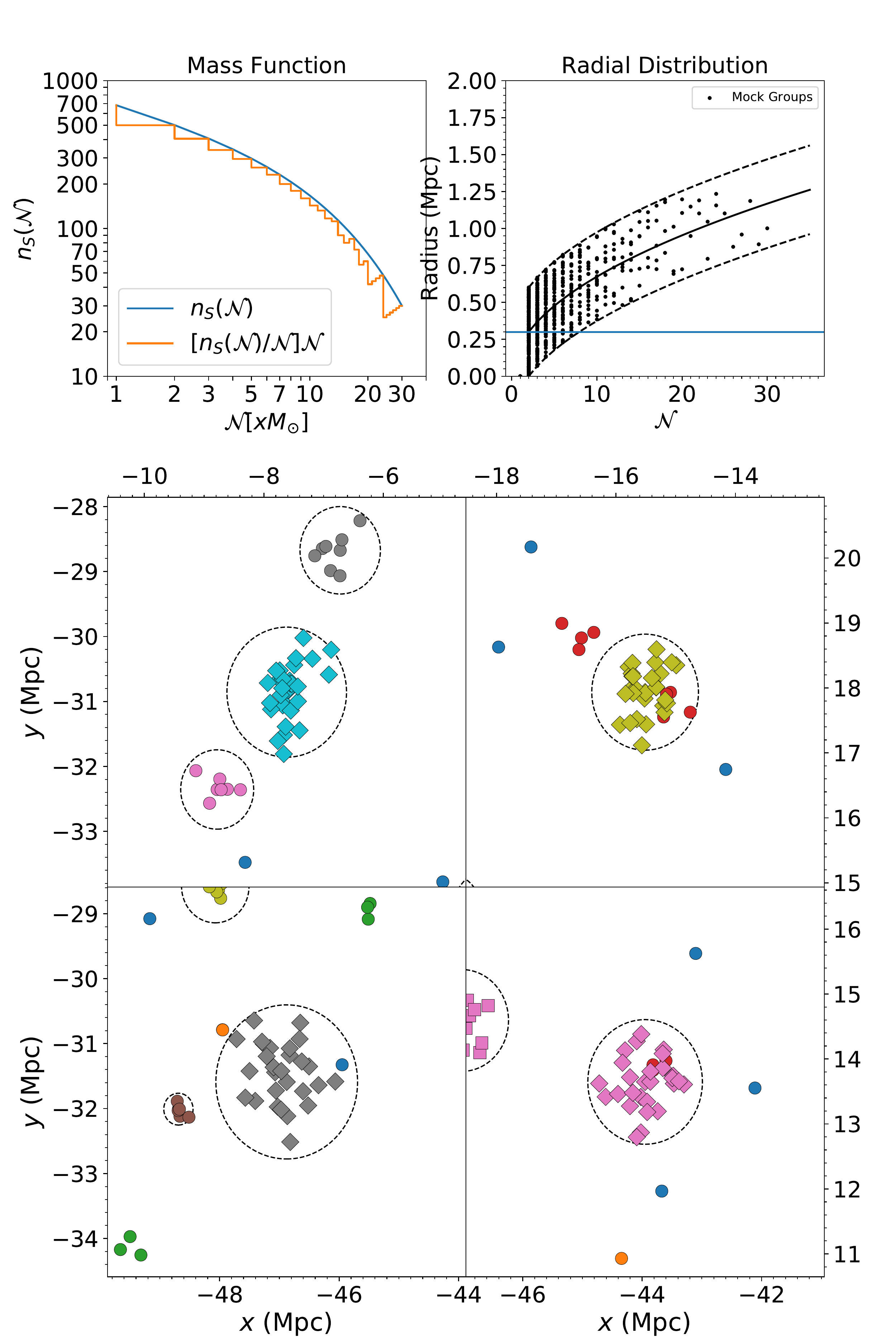}
\caption[Mass distribution of the mock catalogue]{\textbf{Top left:} The mass function which defines our mock catalogue. The blue curve is \cref{eq:schechter} and the orange curve is the final mass distribution. At each value of $\mathcal{N}$, the orange curve is the number of galaxies contained within the maximum integer number of groups with size $\mathcal{N}$ allowed by the blue curve (see text). \textbf{Top right:} Radial size of groups as a function of group membership. The black points are the mock galaxies and are bound by the dashed black lines. The blue horizontal line marks the chosen $\sigma_{\rm{ker}}$: only pairs where the two galaxies are closer than about this value are grouped as pairs by \texttt{TD-ENCLOSER}. \textbf{Bottom:} We show the four largest groups in the mock catalogue. Around each group of five or more members, the maximum radius is shown as a dashed circle. Each group is identified by a colour and marker.}
\label{fig:input_dist}
\end{figure*}
\section{Testing \texttt{TD-ENCLOSER} performance}
\label{sec:test}
\subsection{Parameter sensitivity}
\subsubsection{Mock catalogue}
\texttt{TD-ENCLOSER} has four free parameters compared with their six parameters of \texttt{HOP} (not including $\sigma_{\rm{ker}}$), but unlike other density-based algorithms, our result \textit{does} depend on the order in which particles are considered. We perform similar tests to \citetalias{eisenstein1998hop} to show how our results depend on our choice of parameters. Rather than use a particular distribution of galaxies, we randomly generate a mock galaxy catalogue within a box 100 Mpc across. To make our test as simple as possible, we assume each mock galaxy has the same mass $x$. With this assumption, we use a Schechter function to sample the group membership\footnote{If a group has 30 members where each member is identical and has a mass of $x$, then the group will have a mass of $30x$.} (see the top left panel in \cref{fig:input_dist}). To scale the Schechter function, we choose a maximum group size of 30 members. We arrive at this number by roughly comparing with the group catalogue of \cite{tempel2017groups} (their Table 2). We then define our mass function $n_S(\mathcal{N})$ as:

\begin{equation}
n_S(\mathcal{N})=A10^{(\alpha+1)(\mathcal{N}-\mathcal{N}_0)}e^{-10^{(\mathcal{N}-\mathcal{N}_0)}}
\label{eq:schechter}
\end{equation}

where $\alpha=-1.35$, $\mathcal{N}$ is the group membership and $A$ is a scale factor. We choose our break $\mathcal{N}_0$ to be 15 as it is simply half the maximum allowed group size. We use a scale factor $A \approx 282.5$\footnote{$A=30/0.10618=282.53908$ to five significant figures.} so that $n_S(30)=30$ i.e. a group of 30 members has a mass of 30 in units of $x$. However, the statement $n_S(\mathcal{N})=\mathcal{N}$ is only true for $\mathcal{N}=30$ due to the shape of the function. As we require integer numbers of groups with membership $\mathcal{N}$, we take the number of groups with membership $\mathcal{N}$ as $\floor*{n_S(\mathcal{N})/\mathcal{N}}$ where $\floor*{}$ denotes floor.\par
To populate the box, we generate $(x_{\rm{grp}},y_{\rm{grp}})$ coordinates for each \textit{group} from a random uniform distribution within the limits of the box. For each group with $\mathcal{N}_{\rm{mem}}>1$, we assign a radius $r=\sqrt{\mathcal{N}/7\pi} + f_r$ where $f_r$ is randomly chosen from a uniform distribution with range $[-0.3,0.3]$ (see the top right panel of \cref{fig:input_dist}). We choose 7 as an empirical factor so that roughly 50\% of galaxy pairs are separated by 0.3 Mpc or less (so as to be picked up by \texttt{TD-ENCLOSER}). Clearly this simple prescription assumes that the area of a group is proportional to its membership. For each group, we sample relative $(x_{\rm{gal}},y_{\rm{gal}})$ coordinates for $\mathcal{N}$ galaxies between $[-r,r]$. The final coordinate for a given galaxy is then $(y_{\rm{grp}}+x_{\rm{gal}}, y_{\rm{grp}}+y_{\rm{gal}})$.\par

\begin{figure*}
\centering
\includegraphics[width=0.75\textwidth]{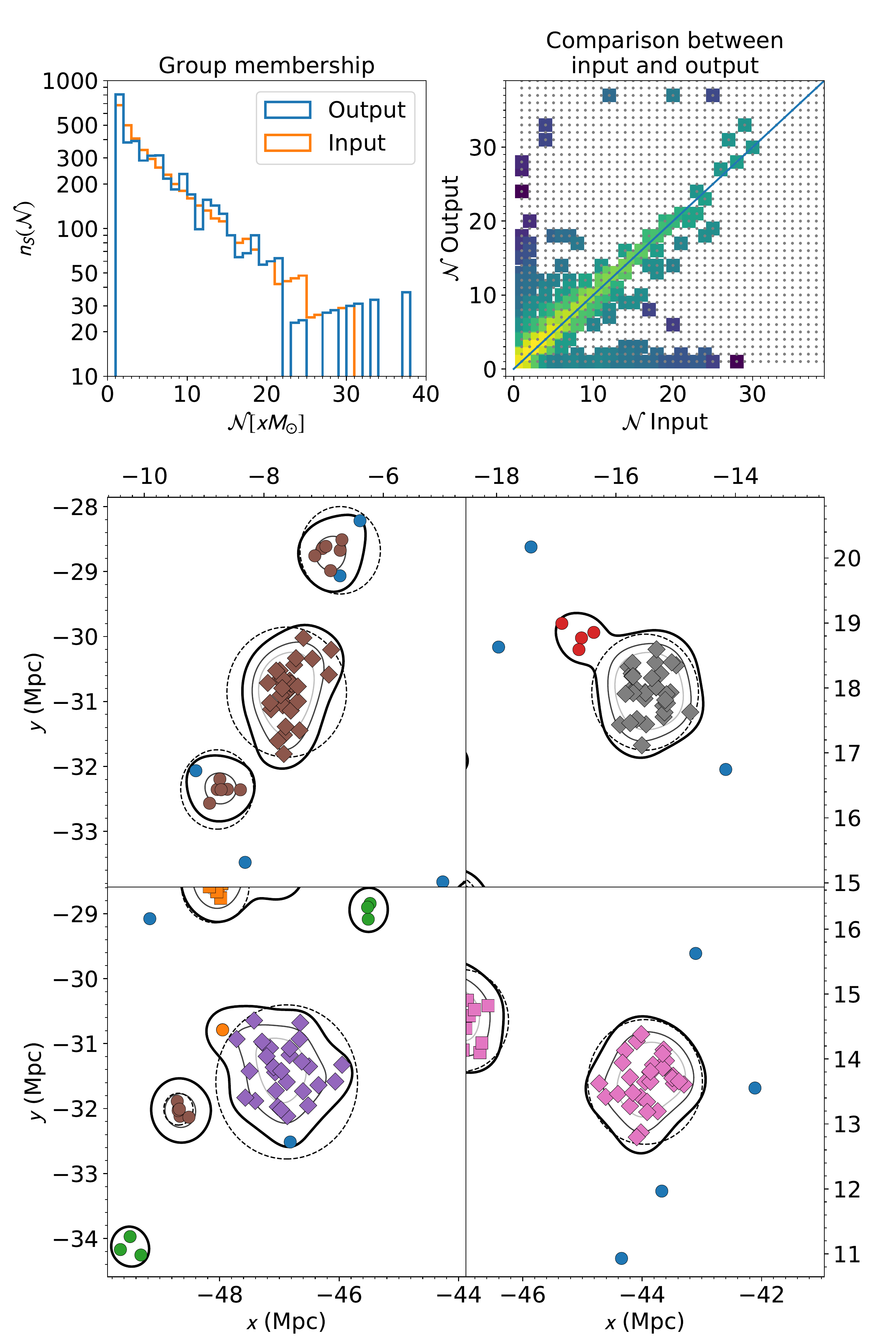}
\caption[Group finder applied to the mock catalogue]{\textbf{Bottom:} The same as the bottom panel of \cref{fig:input_dist} except that galaxies are coloured according to the groups which they have been assigned by \texttt{TD-ENCLOSER}. The contours are $\rho_{\rm{outer}}$ (black), $2.5\rho_{\rm{outer}}=\rho_{\rm{saddle}}$ (dark grey) and $4\rho_{\rm{outer}}$ (light grey). Isolated galaxies are shown as blue circles and galaxy pairs are shown as orange circles. The colours continue in sequence, and the marker indicates the group size in multiples of 10: circles are between 1 and 10, squares are between 11 and 20, diamonds are between 21 and 30 and triangles are between 31 and 40.}
\label{fig:output_dist}
\end{figure*}

In the lower panel of \cref{fig:input_dist}, we show the four largest groups in our mock catalogue. Each group with five or more members is enclosed by a circle with a radius equal to the maximum radius allowed for each group. This panel highlights why we neither want nor need to recover the input distribution with \texttt{TD-ENCLOSER}. Firstly, as the coordinates of both galaxies and groups are randomly chosen, it is likely that some groups will overlap and become a single group, or even some isolated galaxies might lie embedded within another group. \texttt{TD-ENCLOSER} does not know the true input distribution. In fact, we would rather that not all groups are simple isolated circular objects, but actually reflect the real Universe, where larger groups can have substructure. Furthermore, we have programmed \texttt{TD-ENCLOSER} to sensibly clip outliers, and so some galaxies from large groups may be detached from their original groups.\par
Nevertheless, it is still useful to compare the mass function of the input with the mass function determined using the default parameters, even if individual galaxies are not in similar groups in both. As shown in \cref{fig:output_dist}, the ``recovered'' mass function is not too dissimilar from the input mass function. The discrepancies are due to the reasons outlined above. In particular, the mass of individual galaxies is less in the ``recovered'' mass function compared to the input mass function, while the mass in galaxy pairs in greater in the ``recovered'' mass function compared to the input mass function. This is because our radial size prescription was calibrated such that approximately half of all galaxy pairs lie have a separation smaller than 0.3 Mpc (see top right panel of \cref{fig:input_dist}). Hence, we expect fewer galaxy pairs and more isolated galaxies in the ``recovered'' distribution according to our prescription.\par

\subsubsection{Testing parameter sensitivity}
In order to check how the group assignments depend on each parameter, we vary the parameters $\sigma_{\rm{ker}}$, $\rho_{\rm{saddle}}$, $\rho_{\rm{peak}}$ and $N_{\rm{merge}}$ in turn and rerun \texttt{TD-ENCLOSER} on the mock catalogue. We do not perturb the values by a large amount because there are restrictions on each parameter, but also because we already have a good idea about what the default values should be. For example, we have already found $\sigma_{\rm{ker}}=0.3$ Mpc gives the most faithful representation of the true galaxy distribution. We know from \citetalias{eisenstein1998hop} that $\rho_{\rm{outer}}$ does have a significant effect on the result and hence we do not vary this parameter at all. For the other three parameters, there are stricter constraints. $\rho_{\rm{saddle}}$ cannot be equal to $\rho_{\rm{outer}}$ if it is to clip outliers effectively, and in practice, it should be at least $2\rho_{\rm{outer}}$. $\rho_{\rm{peak}}$ must not be equal (or even very close) to $\rho_{\rm{saddle}}$ regardless of the value of $\rho_{\rm{saddle}}$ as this will split groups unnecessarily. We suggest that $\rho_{\rm{peak}} - \rho_{\rm{saddle}} \geq 0.25\rho_{\rm{outer}}$ as a minimum separation. Finally, $N_{\rm{merge}}$ must be greater than or equal to two, as at least two galaxies are required to calculate a mean. In the following test, we keep to these limits so that \texttt{TD-ENCLOSER} can operate as desired. If the results depend only weakly on each parameter, we can be confident in our ability to effectively assign galaxies to groups.\par

\subsubsection{Test results}
The results of our test are shown in \cref{fig:vary_params}, where for each panel, we change one parameter keeping all others the same. To indicate which parameter we are varying, we use the notation [vary$(P)$] where $P$ is the parameter that is \textit{not} set to the default value. If all parameters are set to the default values, then we use the shorthand [default]. Rather than show the mass function as in \cref{fig:output_dist}, we take the \textit{difference} between the group richness $\mathcal{N}$ determined using the default parameters ($\mathcal{N}^{[\rm{default}]}$) and the perturbed parameters ($\mathcal{N}^{[\rm{vary}(P)]}$) for each mock galaxy, and compute the histogram for all mock galaxies. This is \textit{not} the same as comparing the \textit{in}put from \cref{fig:input_dist} with the \textit{out}put from \cref{fig:output_dist}, as here we are comparing the \textit{out}put using different values for the free parameters. To give a complete picture, we could replace the $x$ and $y$ axis in the top-right panel of \cref{fig:output_dist} with $\mathcal{N}^{[\rm{default}]}$ and $\mathcal{N}^{[\rm{vary}(P)]}$ respectively. However, this would give us eight panels to present, two for each parameter. Instead, we show the difference in the richness distribution with varying parameters. The result is \cref{fig:vary_params}, where each panel corresponds to a parameter, and each panel contains two histograms corresponding to each value we choose.\par
We go through each panel of \cref{fig:vary_params} in turn. Each histogram can be interpreted in the following way: if a single galaxy is near a group of 20 galaxies when $\sigma_{\rm{ker}}=0.3$, then when $\sigma_{\rm{ker}}=0.4$, it becomes part of a new group of 21 members. In this case, $\mathcal{N}^{[\rm{vary}(\sigma_{\rm{ker}})]}=\mathcal{N}^{[\rm{default}]}+1$ for 20 galaxies, and $\mathcal{N}^{[\rm{vary}(\sigma_{\rm{ker}})]}=\mathcal{N}^{[\rm{default}]}+20$ for one galaxy.\par

\subsubsection{Varying $\sigma_{\rm{ker}}$}
We select $\sigma_{\rm{ker}}=0.2$ and 0.4 Mpc. As expected, setting $\sigma_{\rm{ker}}=0.2$ Mpc generally results in galaxies forming smaller groups in the first pass compared to $\sigma_{\rm{ker}}=0.3$ Mpc. A small fraction of galaxies ($\sim 1 \%$) join larger groups due to the fact that by using a smaller kernel, some groups that would have reached $\rho_{\rm{peak}}$ when $\sigma_{\rm{ker}}=0.3$ would not have reached $\rho_{\rm{peak}}$ when $\sigma_{\rm{ker}}=0.2$. Hence, these groups would not have been clipped in the second pass with $\sigma_{\rm{ker}}=0.2$ and therfore the galaxies which were clipped when $\sigma_{\rm{ker}}=0.3$ have appeared to join a larger group, even though the kernel size has decreased. However, this occurrence is rare because $\rho_{\rm{peak}}$ is close to $\rho_{\rm{saddle}}$. Conversely, setting $\sigma_{\rm{ker}}=0.4$ Mpc results in more galaxies residing in larger groups compared to the default choice. Here, about $2 \%$ of galaxies join smaller groups because they are clipped when $\sigma_{\rm{ker}}=0.4$ but not when $\sigma_{\rm{ker}}=0.3$, and hence belong to a smaller group even though the kernel size has increased. The result is most sensitive to this parameter out of the four parameters as evidenced by the lower height of each peak at $\mathcal{N}^{[\rm{vary}(\sigma_{\rm{ker}})]}-\mathcal{N}^{[\rm{default}]}=0$ (indicating no change) and the broad distribution. For $\sigma_{\rm{ker}}=0.2$ and 0.4 Mpc, 95\% of galaxies change group membership by less than 15 and 9 members respectively.

\subsubsection{Varying $\rho_{\rm{saddle}}$}
We select $\rho_{\rm{saddle}}=3.6$ and $4.4$. A value of $\rho_{\rm{saddle}}=3.6$ tends to increase group sizes (and reduce the number of groups) as some galaxies are less likely to be clipped in the second pass than when $\rho_{\rm{saddle}}=4$. This makes sense conceptually as all groups above $\rho_{\rm{peak}}$ essentially grow and absorb nearby groups. A very small fraction of galaxies move to smaller groups which occurs when a small group is split up and one part joins a nearby large group and the other part then forms a small group. In total, 95\% of galaxies change group membership by less than 3. Choosing a value of $\rho_{\rm{saddle}}=4.4$ means that galaxies are more likely to be clipped from their original groups. Again, 95\% of galaxies change group membership by less than 3.

\subsubsection{Varying $\rho_{\rm{peak}}$}
\begin{figure*}
\centering
\includegraphics[width=0.7\textwidth]{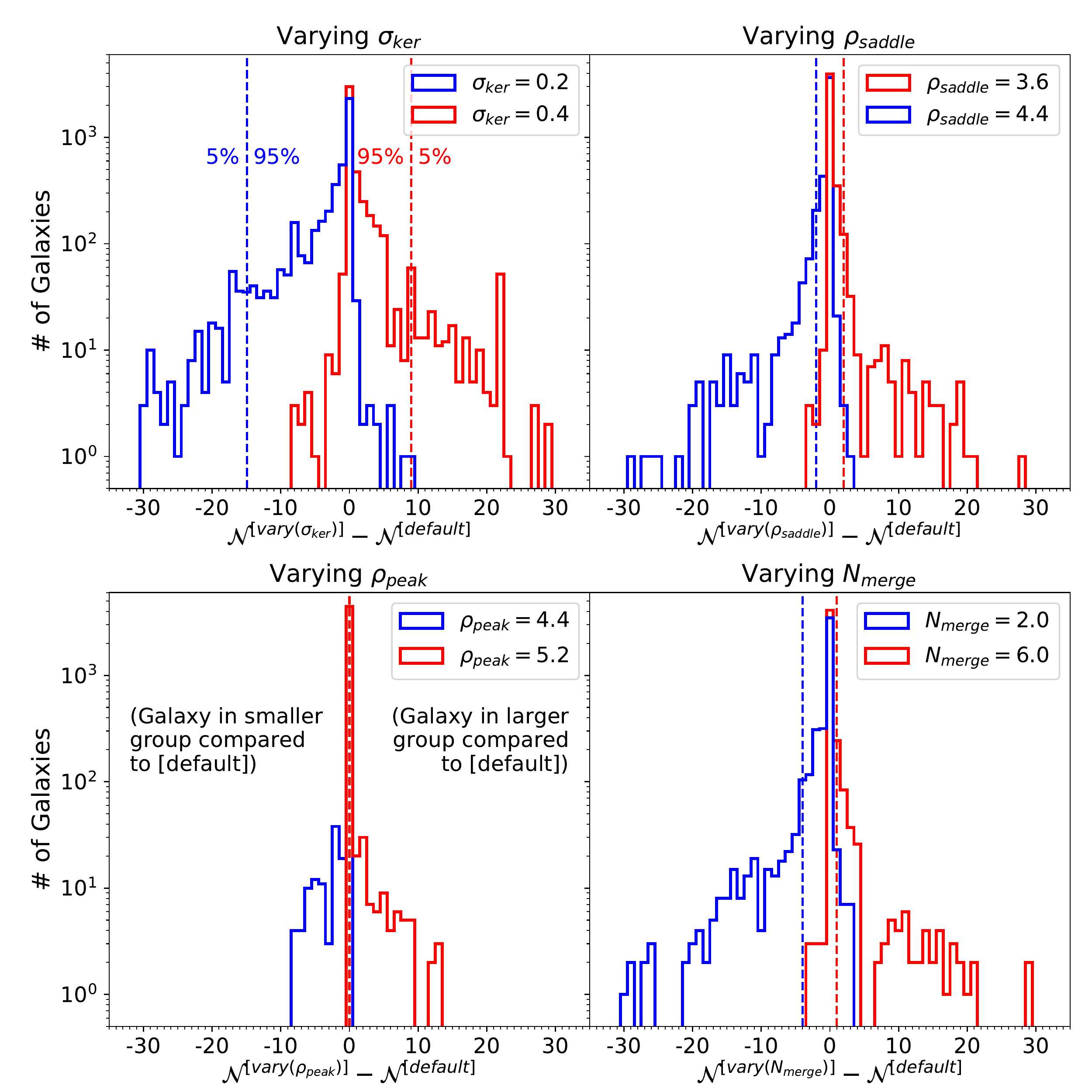}
\caption[Group sensitivity to free parameters]{Histograms illustrating how the group membership depends on the free parameters of \texttt{TD-ENCLOSER}. In each panel, a single parameter is changed from the default value to two different values either side of the default, corresponding to the two histograms in each panel. In each panel, the $x$-axis is the \textit{difference} between the richness of the group a particular \textit{galaxy} belongs to using the \textit{perturbed} parameters, and the richness of the group using the default parameters.}
\label{fig:vary_params}
\end{figure*}

We select $\rho_{\rm{peak}}=4.4$ and 5.2. As seen in the third panel of \cref{fig:vary_params}, this is the parameter that the group membership is the least sensitive to. This is because varying $\rho_{\rm{peak}}$ only changes \textit{which} groups are clipped in the second pass and doesn't have any bearing on how \textit{much} those groups are clipped. By lowering $\rho_{\rm{peak}}$, more groups are clipped. If $\rho_{\rm{peak}}$ is increased compared to the default value, then fewer groups are clipped. In total, 95\% of galaxies do not change group membership at all.

\subsubsection{Varying $N_{\rm{merge}}$}
We select $N_{\rm{merge}}=2$ and 8. Using $N_{\rm{merge}}=2$, galaxies at the fringe of $\rho_{\rm{saddle}}$ are more likely to be clipped during the second pass compared with $N_{\rm{merge}}=4$. Hence, groups are more likely to be split up into a large subgroup and one or more smaller subgroups (extended blue tail in the fourth panel). Using $N_{\rm{merge}}=6$ means that galaxies are less likely to be clipped (extended red tail in the fourth panel). Because this parameter is only relevant in the second pass, and then only when a group reaches $\rho_{\rm{peak}}$, the sensitivity of the result on $N_{\rm{merge}}$ is less than that of $\sigma_{\rm{ker}}$ and $\rho_{\rm{saddle}}$, and only slightly more than $\rho_{\rm{peak}}$. This can be seen as 95\% of galaxies change membership by 4 or less when $N_{\rm{merge}}=2$, and by 1 or zero when $N_{\rm{merge}}=6$.

\begin{figure}
\centering
\includegraphics[width=0.49\textwidth]{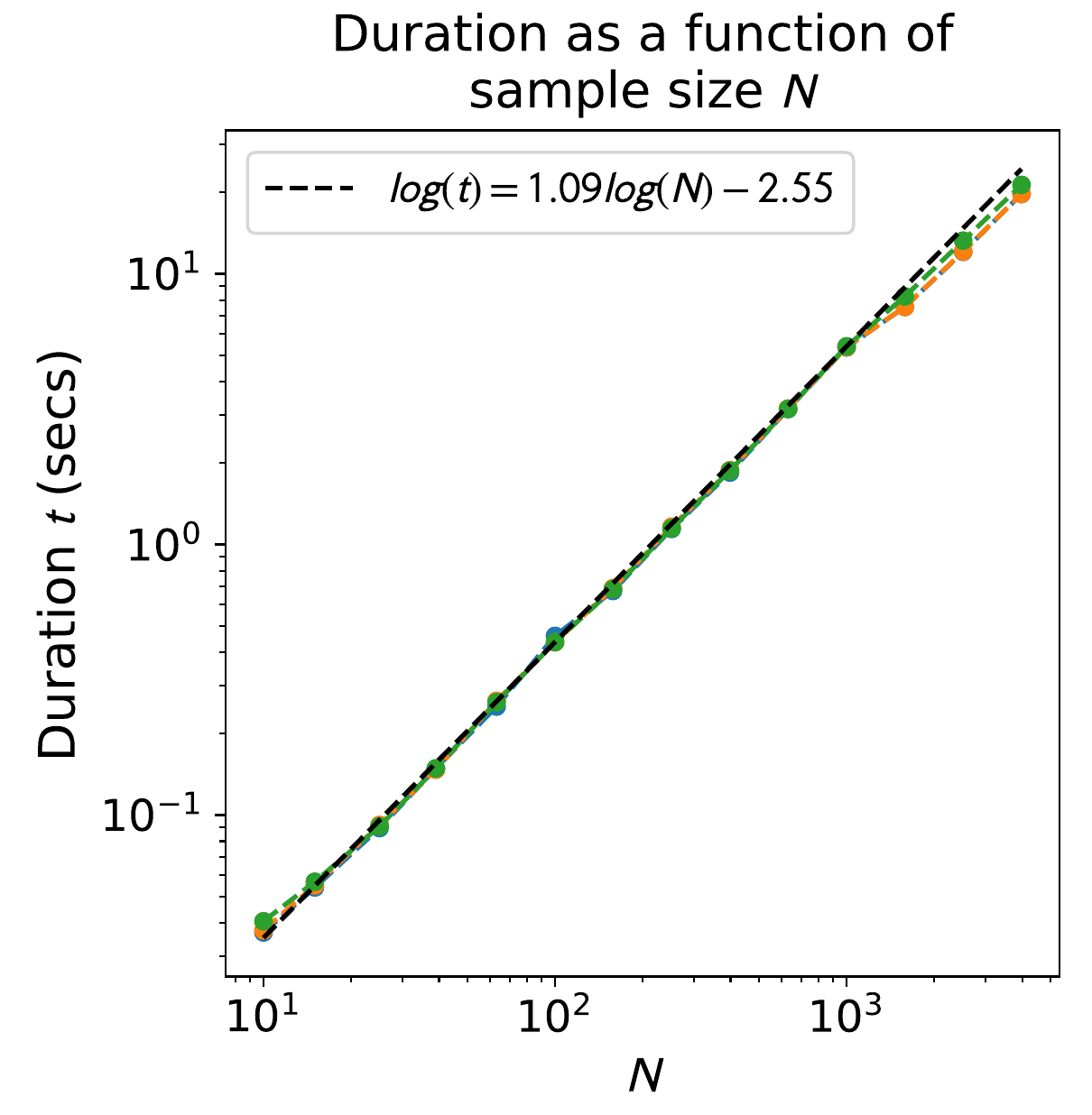}
\caption[\texttt{TD-ENCLOSER} running time]{Duration of a \texttt{TD-ENCLOSER} run as a function of sample size. Three different runs are shown in orange, blue and green. The dashed line is the connecting line between the points at $N=100$ and $N=1000$ and is given in the legend.} 
\label{fig:group_speed}
\end{figure}

\subsection{\texttt{TD-ENCLOSER} running speed}
Finally, in \cref{fig:group_speed}, we assess the speed of \texttt{TD-ENCLOSER}. As \texttt{TD-ENCLOSER} is designed to only consider the local environment close to a ``target'' galaxy, it is not optimised for large sample sizes. We find that the elapsed time depends linearly on the sample size, taking approximately 5 seconds to iterate through 1000 galaxies using a late 2013 iMac computer. We check to see if the speed varies with different choices in the parameters (not shown). We find that for $N\gtrsim100$, choosing parameters that reduce the number of galaxies to consider in the second and third passes decreases the performance time. Of these, choosing a lower clipping threshold (i.e. $\rho_{\rm{saddle}}$) gives the largest reduction in speed because fewer galaxies are clipped. Choosing a larger $N_{\rm{merge}}$ also reduces the number of galaxies to be clipped. Finally, by choosing a higher $\rho_{\rm{peak}}$, fewer groups are clipped in the second pass, although the improvement is minor overall.\par
Choosing values that increases the number of galaxies to be considered after the first pass increases the time taken. The most significant increase is seen with $\rho_{\rm{saddle}}$, where an increase of 10\% in $\rho_{\rm{saddle}}$ results in a 10\% increase in calculation time. Decreasing $N_{\rm{merge}}$ by a factor of two compared to the default value only results in a 5\% increase in performance time, and there is almost no change with $\rho_{\rm{peak}}$. $\rho_{\rm{saddle}}$ produces the biggest changes overall to performance time.\par
\section{Accounting for group multiplicity}
\label{sec:multiplicity}
As discussed at the end of \cref{sec:intro2}, we will likely find many duplicates of the same intrinsic group when we run \texttt{TD-ENCLOSER} on a set of neighbouring galaxies that changes depending on which MaNGA galaxy is the target galaxy. Suppose that a given galaxy group contains four MaNGA galaxies, [$M_1$, $M_2$, $M_3$, $M_4$], where each MaNGA galaxy (e.g. $M_1$) plays host to its own set of neighbours (e.g. $S_1$). Hence, we will find four sets, one for each MaNGA galaxy. Of course, we did not know beforehand that this group would be found four times, although we could have predicted it by finding the four MaNGA galaxies in advance. In the case that they are identical, we can randomly select one of the groups to represent all four. However, we may be presented with a case where not all sets are equal. In fact, it is possible that not all four MaNGA galaxies will appear in all four sets. In another example, a MaNGA galaxy can appear on the outskirts of a large cluster with hundreds of member galaxies, but can also appear as a small group. In these cases, we choose a subset of representative set to represent the intrinsic groups so that each MaNGA galaxy appears only once in our catalogue.  
We achieve this aim using the following steps:
\begin{enumerate}
\item We select $N_1$ MaNGA galaxies which are contained within set $S_1$. 
\item We select $N_2$ sets that contain at least one of the MaNGA galaxies belonging to set $S_1$. $N_2$ may be larger than $N_1$.
\item We select $N_3$ MaNGA galaxies that are found in the $N_2$ set identified in step 2. 
\item If $N_3>N_1$, then we repeat steps 1. - 3. selecting the set of $N_3$ MaNGA galaxies which are found in set $N_2$. If $N_3=N_1$, we move onto step 5.
\item We select all sets that are larger than half the maximum richness, $\mathcal{N}_{\rm{max}}$, of the final sets. This step ensures we are likely to select duplicates or variations of the same large group, deselecting nearby small groups and isolated galaxies that are otherwise connected to the large group.
\item Of the subset of sets that satisfy $\mathcal{N} > \mathcal{N}_{\rm{max}}/2$, we calculate the median redshift of the host galaxies in the subset. 
\item We select the set $S_M$ which encloses/is hosted by the MaNGA galaxy with the median redshift and assign this set to the final selection. If there are more than one MaNGA galaxies at the median redshift, we randomly select one.
\item We remove \textit{all} sets which contain the same \textit{MaNGA} galaxies as set $S_M$ from our final selection.
\item If there are remaining sets that do not share any MaNGA galaxies in common with set $S_M$, we randomly \textit{non-unique} galaxies within the remaining sets\footnote{Consider four sets: [$M_1,M_2,M_3$], [$M_1$], [$M_2$] and [$M_3$]. If we randomly select by set, then the set containing three MaNGA galaxies has only a 25\% chance of being selected. If we choose by MaNGA galaxy and select its enclosing set, then the set containing three MaNGA galaxies has a 50\% chance of being selected. We choose to select using the latter method in step 9.}, and repeat step 8.
\item We repeat step 9. until all MaNGA galaxies are assigned to one set only.
\end{enumerate}

We illustrate this selector algorithm in \cref{fig:groupselector}. We start with set $S_1$ and find that it hosts five MaNGA galaxies (step 1.). We then check all other sets and find five more sets that contain at least one of these five MaNGA galaxies (step 2.). Two of these (sets $S_5$ and $S_6$) contain MaNGA galaxies that are not found in set $S_1$, and so we search for all sets that contain these new MaNGA galaxies (step 3.). We find four new sets that do not contain any of the original five MaNGA galaxies, but are linked to set $S_1$ by a chain of MaNGA galaxies (step 4.). As we do not find any more sets containing these new MaNGA galaxies, the tree stops growing here.\par
Following step 4. above, we select all sets that are richer than half the maximum richness. In this example, set $S_5$ is the richest set with 56 members, and so we select sets $S_1$, $S_5$, $S_6$, $S_8$ and $S_{10}$. Of these five sets, we select the set with the median redshift, which in this case is set $S_{10}$ (steps 6. and 7.). This is the first set to make the final selection. We deselect all sets that share at least one MaNGA galaxy in common with set $S_{10}$, namely sets $S_5$, $S_7$ and $S_8$ (step 8.). We randomly select one of the remaining sets, which happens to be set $S_1$ (step 9.). The only remaining set that doesn't share any MaNGA galaxies in common with set $S_1$ is set $S_9$, and so the final selection contains sets $S_1$, $S_9$ and $S_{10}$ (step 10.). We could also select sets $S_2$, $S_3$, $S_4$, $S_6$, $S_9$ in step 9., in which case the final selection would be slightly different.\par
There are a few alternative ways we could use to select the representative set(s). For example, we could take the largest set found as that is most likely to contain all of the MaNGA galaxies. However, that may bias us towards larger groups. We could also randomly select the representative set from all possible sets, but that may select a small set instead of a larger set just by chance. For a large \textit{group} with many MaNGA galaxies, we may miss a substantial fraction of the group if we were to select MaNGA galaxies at the extreme velocities. Therefore, by choosing a set based on the median redshift of the MaNGA galaxies, we expect the cylinder to encompass the ``true'' velocity extent of the group.\par
It is possible that some MaNGA galaxies won't make it into the final selection, depending on the exact galaxy configuration in three dimensional space. If we take our earlier example of four MaNGA galaxies, but split them into three sets of [$M_1$, $M_2$, $M_3$] and one set of [$M_2$, $M_3$, $M_4$], then according to our random selection, $M_4$ is likely to be excluded. We find that out of nearly 4600 galaxies, about 3\% don't make it into the final selection. These galaxies will essentially only live in small groups, and so we do not expect any bias to come from this.

\tikzstyle{block} = 
    [
        rectangle
      , draw
      , text width=6.5em
      , text centered
      , rounded corners
      , minimum height=2em
      ]

\tikzstyle{line} = 
    [
        draw
     , -latex'
     ]
     
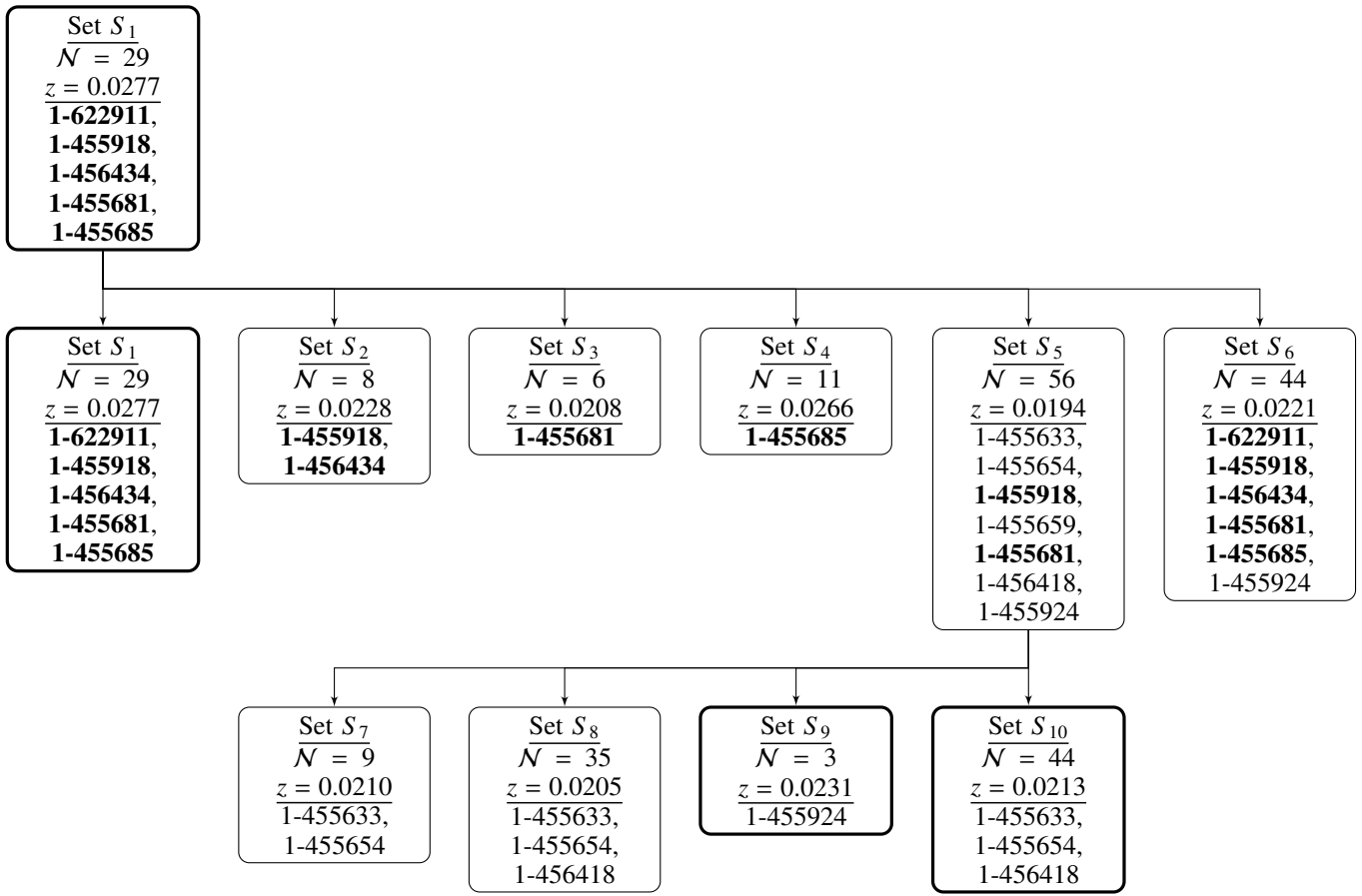
\begin{figure*}
\centering
 \begin{tikzpicture}       
     \matrix (mat) [matrix of nodes, nodes=block, column sep=5mm, row sep=1cm] 
{
&\node [block, very thick] (1) {\underline{Set $S_1$}\\ $\mathcal{N} = 29$ \underline{$z=0.0277$} \textbf{1-622911}, \textbf{1-455918}, \textbf{1-456434}, \textbf{1-455681}, \textbf{1-455685}};  & & & & &\\ 
&\node [block, very thick] (2) {\underline{Set $S_1$}\\ $\mathcal{N} = 29$ \underline{$z=0.0277$} \textbf{1-622911}, \textbf{1-455918}, \textbf{1-456434}, \textbf{1-455681}, \textbf{1-455685}};  
&\node [block] (3) {\underline{Set $S_2$}\\ $\mathcal{N} = 8$ \underline{$z=0.0228$} \textbf{1-455918}, \textbf{1-456434}};  
&\node [block] (4) {\underline{Set $S_3$}\\ $\mathcal{N} = 6$ \underline{$z=0.0208$} \textbf{1-455681}};  
&\node [block] (5) {\underline{Set $S_4$}\\ $\mathcal{N} = 11$ \underline{$z=0.0266$} \textbf{1-455685}};  
&\node [block] (6) {\underline{Set $S_5$}\\ $\mathcal{N} = 56$ \underline{$z=0.0194$} 1-455633, 1-455654, \textbf{1-455918}, 1-455659, \textbf{1-455681}, 1-456418, 1-455924}; 
& \node [block] (7) {\underline{Set $S_6$}\\ $\mathcal{N} = 44$ \underline{$z=0.0221$} \textbf{1-622911}, \textbf{1-455918}, \textbf{1-456434}, \textbf{1-455681}, \textbf{1-455685}, 1-455924};     \\
& &\node [block] (8) {\underline{Set $S_7$}\\ $\mathcal{N} = 9$ \underline{$z=0.0210$} 1-455633, 1-455654};  
& \node [block] (9) {\underline{Set $S_8$}\\ $\mathcal{N} = 35$ \underline{$z=0.0205$} 1-455633, 1-455654, 1-456418};  
&\node [block, very thick] (10) {\underline{Set $S_9$}\\ $\mathcal{N} = 3$ \underline{$z=0.0231$} 1-455924}; 
& \node [block, very thick] (11) {\underline{Set $S_{10}$}\\ $\mathcal{N} = 44$ \underline{$z=0.0213$} 1-455633, 1-455654, 1-456418};  &   \\
};  

\path[line] (1.south)    --+(0,-0.5) -| node [pos=0.3, above] {} (2.north);
\path[line] (1.south)   -- +(0,-0.5) -| node [pos=0.3, above] {} (3.north);
\path[line] (1.south)   -- +(0,-0.5) -| node [pos=0.3, above] {} (4.north);
\path[line] (1.south)   -- +(0,-0.5) -| node [pos=0.3, above] {} (5.north);
\path[line] (1.south)   -- +(0,-0.5) -| node [pos=0.3, above] {} (6.north);
\path[line] (1.south)   -- +(0,-0.5) -| node [pos=0.3, above] {} (7.north);
\path[line] (6.south)    --+(0,-0.5) -| node [pos=0.3, above] {} (8.north);
\path[line] (6.south)   -- +(0,-0.5) -| node [pos=0.3, above] {} (9.north);
\path[line] (6.south)   -- +(0,-0.5) -| node [pos=0.3, above] {} (10.north);
\path[line] (6.south)   -- +(0,-0.5) -| node [pos=0.3, above] {} (11.north);
\end{tikzpicture}
    \caption[Finding unique groups]{\textbf{Finding unique groups.} Here we show an example where 10 galaxy sets are connected by a network of MaNGA galaxies. For each set, we give the richness $\mathcal{N}$ and the redshift of the ``host'' MaNGA galaxy. The starting set is set $S_1$, and its member galaxies are emphasised in bold. The final representative sets selected by the selector are set $S_1$, $S_9$ and $S_{10}$ which are indicated by thick borders. These three sets do not share any MaNGA galaxies but may share galaxies not in MaNGA. This example has many possible solutions (see the text for details).}
    \label{fig:groupselector}
\end{figure*}


\section{Conclusions}
In \cref{sec:td-encloser}, we introduced a new group finder algorithm (\texttt{TD-ENCLOSER}) which has some features in common with the HOP method of EH96 (see \cref{fig:group_finder}), but is based on an entirely different method. Its main function is to assign galaxies to regions of high density, before clipping outliers and forming new groups from those outliers (\cref{fig:groupfinder_1d}). \texttt{TD-ENCLOSER} is different to most other group-finder algorithms in that it is used to discover which group encloses a particular galaxy of interest. It is not designed to produce large group catalogues of hundreds of thousands of galaxies but can be used to obtain the local galaxy distribution. It works on the simple assumption that the gradient along a one-dimensional straight line between two points encodes the local two-dimensional topology. If the gradient of the connecting line between a galaxy and a nearby peak does not fall below a small noise threshold $\epsilon$, and if the galaxy is in a sufficiently dense environment, then we assign the galaxy to that peak. If the galaxy satisfies the threshold to be in a group but is sufficiently far enough from the peak to be near the outskirts, it is ejected from the peak after which it seeks to join a new, smaller group. As with any algorithm, there will inevitably be anomalies, especially when the contour topology is complex. However, \texttt{TD-ENCLOSER} has already been used on thousands of real life cases (see \citetalias{graham2019bcatalogue} for an overview and \cite{graham2019dclusters} (Paper IV) for a few key examples) and has been thoroughly checked for reliability.\par
In \cref{sec:test}, we tested \texttt{TD-ENCLOSER} on a mock catalogue of galaxies using a Schechter mass function \citep{schechter1976function} to define the group membership (\cref{fig:input_dist}). While we did not aim to reproduce exactly the input distribution, we found that we could match the input mass function with reasonable accuracy (top left panel of \cref{fig:output_dist}). The reason for the discrepancy is that many pairs are broken up (by design) and a small percentage of individual galaxies can dramatically change their group membership (top right panel of \cref{fig:output_dist}). We found that of the four parameters where we have some freedom to choose the default values, \texttt{TD-ENCLOSER} is reasonably insensitive to all but one of them (\cref{fig:vary_params}). Again, individual galaxies can change group membership dramatically depending on the choice of parameters, but the fraction of galaxies with large changes in group membership is small. The highest sensitivity is towards $\sigma_{\rm{ker}}$, which is a known feature of KDE methods. We checked the speed of operation and found that it increases linearly with the sample size (\cref{fig:group_speed}). The parameter that introduces the most variation is $\rho_{\rm{saddle}}$ where a variation of 10\% results in a similar change in performance time.\par
In preparation for the task of constructing the group catalogue in \citetalias{graham2019bcatalogue}, we have developed a simple procedure to select one or more representative groups of a set of duplicates linked by MaNGA galaxies (see \cref{sec:multiplicity}). We select representative sets using knowledge about the group size while incorporating a random selection to ensure we are not biased towards any particular sets.


\section*{Acknowledgements}
Funding for the Sloan Digital Sky Survey IV has been provided by the Alfred P. Sloan Foundation, the U.S. Department of Energy Office of Science, and the Participating Institutions. SDSS acknowledges support and resources from the Center for High-Performance Computing at the University of Utah. The SDSS website is www.sdss.org.\par
SDSS is managed by the Astrophysical Research Consortium for the Participating Institutions of the SDSS Collaboration including the Brazilian Participation Group, the Carnegie Institution for Science, Carnegie Mellon University, the Chilean Participation Group, the French Participation Group, Harvard-Smithsonian Center for Astrophysics, Instituto de Astrofísica de Canarias, The Johns Hopkins University, Kavli Institute for the Physics and Mathematics of the Universe (IPMU) / University of Tokyo, Lawrence Berkeley National Laboratory, Leibniz Institut für Astrophysik Potsdam (AIP), Max-Planck-Institut für Astronomie (MPIA Heidelberg), Max-Planck-Institut für Astrophysik (MPA Garching), Max-Planck-Institut für Extraterrestrische Physik (MPE), National Astronomical Observatories of China, New Mexico State University, New York University, University of Notre Dame, Observatório Nacional / MCTI, The Ohio State University, Pennsylvania State University, Shanghai Astronomical Observatory, United Kingdom Participation Group, Universidad Nacional Autónoma de México, University of Arizona, University of Colorado Boulder, University of Oxford, University of Portsmouth, University of Utah, University of Virginia, University of Washington, University of Wisconsin, Vanderbilt University, and Yale University.\par
This publication makes use of data products from the Two Micron All Sky Survey, which is a joint project of the University of Massachusetts and the Infrared Processing and Analysis Center/California Institute of Technology, funded by the National Aeronautics and Space Administration and the National Science Foundation.





\bibliographystyle{mnras}
\bibliography{MasterBibliography} 





\end{document}